\documentclass{article}
\usepackage{graphicx,amsmath,amssymb,amsfonts}
\usepackage{wrapfig}
\newcommand{\bu}{\mathbf{u}}

\newcommand{\bv}{\mathbf{v}}
\newcommand{\bw}{\mathbf{w}}

\newcommand{\bx}{\mathbf{x}}

\newcommand{\bg}{\mathbf{g}}
\newcommand{\bn}{\mathbf{n}}

\newcommand{\bs}{\mathbf{s}}
\newcommand{\blf}{\mathbf{f}}
\newcommand{\rd}{\mathrm{d}}

\newcommand{\QQ} {\mathbb{Q}}
\newcommand{\VV} {\mathbb{V}}

\newcommand{\dO} {{\partial\Omega}}

\newcommand{\Div}{\operatorname{div}}

\def\dO{\partial\Omega }
\def\cE{{\cal E}}
\newtheorem{theorem}{Theorem}[section]
\newtheorem{remark}{Remark}[section]

\newenvironment{changemargin}[2]{%
\begin{list}{}{%
\setlength{\topsep}{0pt}%
\setlength{\leftmargin}{#1}%
\setlength{\rightmargin}{#2}%
\setlength{\listparindent}{\parindent}%
\setlength{\itemindent}{\parindent}%
\setlength{\parsep}{\parskip}%
}%
\item[]}{\end{list}}

\newenvironment{keywords}{%
\begin{changemargin}{1cm}{1cm}
\noindent{\small\bf Keywords:}
}
{\end{changemargin} }

\title{
A finite element solver and energy stable coupling for 3D and 1D fluid models
\thanks{This work has been supported in part by RFBR grants 12-01-00283, 11-01-00971, 11-01-00767, 12-01-33084}}
\author{Tatiana K. Dobroserdova\thanks{Department of Mechanics and Mathematics, Moscow State University, Moscow; {\tt DobroserdovaTK@gmail.com}} \and
       Maxim A. Olshanskii\thanks{Department of Mathematics, University of Houston, Houston, Texas 77204-3008 and Dept. Mechanics and Mathematics, Moscow State University, Moscow 119899;
{\tt molshan@math.uh.edu}}}

\date{}

\begin{document}

\maketitle

\begin{abstract} The paper develops a solver based on a conforming finite element method
for a 3D--1D coupled incompressible flow problem.  New coupling conditions are introduced to ensure
a suitable bound for the cumulative energy of the model.
We study the stability and  accuracy of the discretization method, and the performance of some state-of-the-art linear algebraic solvers for such flow configurations. Motivated by the simulation of the flow over  inferior vena cava (IVC) filter, we consider the coupling of a 1D fluid model and a 3D fluid model posed in a domain with anisotropic inclusions.
The relevance of our approach to realistic cardiovascular simulations is demonstrated by computing a blood flow over
a model IVC filter.
\end{abstract}


\section{Introduction}
Coupling a 3D fluid flow model and a system of hyperbolic equations posed on a 1D graph is a well
established approach for numerical simulations of blood flows in a system of vessels~\cite{QTV}.
Such a geometric multiscale strategy  is particularly efficient, when the attention to local flow details and the qualitative assessment of global flow statistics are both important. The relevance to cardiovascular simulations and
challenging mathematical problems of coupling parabolic 3D and hyperbolic 1D equations put 3D--1D flow problems
in the focus of intensive research.   Thus, the coupling of a 3D fluid/structure interaction problem with a reduced 1D model merged to outflow boundary, which acts as an absorbing device, was studied in \cite{3D1D1}.
The coupling of a 3D fluid problem with multiple downstream 1D models in the context of a finite element
method was considered in \cite{Vignon}. In \cite{UBVF}, a system of a 3D fluid/structure interaction problem and a 1D finite element method model of the whole arterial tree was implemented to model the carotid artery blood flow; and in \cite{BFU} a unified variational formulation for   multidimensional  models was introduced. A splitting method, extending the  pressure-correction scheme to 3D--1D coupled systems, was studied in \cite{Papa}.

In most of these studies, the 3D model was a generic fluid-elasticity or rigid fluid model, while numerical validations
were commonly  done for cylindric type 3D domains (with rigid or elastic walls); several authors considered geometries with bifurcation~\cite{Vignon,UBVF} or constrained geometries (modeling  a stenosed artery) \cite{BFU}.
More complicated  geometries occur in simulations of blood flows, if one is interested in modeling the effect of endovascular implants, such as inferior vena cava (IVC) filters. In numerical  simulations, a part of a vessel  with an intravenous filter leads  to the computational 3D domain with strongly anisotropic inclusions. A downstream flow behind the implant may exhibit a complex structure with traveling vortices,  swirls, and recirculation regions (the latter may occur if plaque  is captured by the filter). Moreover, the IVC flow is strongly influenced by the contraction of the heart, and both forward (towards the heart) and reverse (from the heart) flows occur within one cardiac cycle. Downstream coupling  conditions for such flows may be a delicate issue. Thus, the flow over an IVC filter is an interesting and challenging  problem for a 3D-1D flow numerical solver.

The coupling conditions of 3D and 1D fluid models and their properties were studied by several authors.
Coupling conditions and algorithms based on subdomain iterations were introduced in  \cite{3D1D1}, and
the stability properties of each subproblem were analyzed separately.  The first analysis of two models together was done in  \cite{3D1D2}. In that paper, it was noted that if  the Navier-Stokes equations   are taken in the rotation form and suitably  coupled with a 1D downstream flow model, then one can show a bound for the  joint energy of the system. It is, however, well known that using a finite element method for the rotation form of the   Navier-Stokes equations needs special care~\cite{LMNOR}, and setting appropriate outflow boundary conditions can be an issue. In the present paper, we introduce an \textit{energy consistent} coupling with a
1D model for the convection form of the Navier-Stokes equations. The joint energy of a coupled 3D-1D model
is appropriately  balanced and dissipates for viscous flows.

Handling highly anisotropic structures  is a well-known
challenge in numerical flow simulations and analysis. There are only a few computational studies
addressing the dynamics of  blood flows in vessels with implanted filters.
Recently, Vassilevski et al.~\cite{VSK} numerically approached the problem of intravenous filter optimization using a finite-difference method on octree cartesian meshes to resolve the geometry of implants. In that paper, it was also discussed how the effect of an implant can be accounted in a 1D model through a modification of a vessel wall state equation (see also \cite{all2,all3} for the development of this method for atherosclerotic blood
vessels).
In the present paper, we take another approach and locally resolve the full 3D model, while keeping the state equation unchanged. We report on a finite element method for modeling a 3D-1D coupled fluid problem, when the 3D domain has anisotropic inclusions. Naturally, this leads  to meshes containing possibly  anisotropic tetrahedra.
We study the performance of the finite element method both by considering the
accuracy of solutions and by monitoring the convergence of one state-of-the-art linear algebra solver for
the systems of linear algebraic equations to be solved  on every time step of the
method. We are interested in the ability of the solver to predict such important
statistics as the drag force experienced by an intravenous implant.

The remainder of the paper is organized as follows. In section~\ref{s_model}, we review  3D and 1D fluid models and discuss coupling conditions. The stability properties of the coupled model are also addressed in section~\ref{s_model}.
Section~\ref{s_solver} presents a time-stepping numerical scheme and an algebraic solver. In section~\ref{s_validate}, we validate the 3D finite element solver
and the coupled method by considering the benchmark problem of a flow past a 3D cylinder and a problem with
an analytical solution. The application of the method to simulate a blood flow over a model IVC filter
is given in section~\ref{s_filter}.  Numerical experiments were performed using the Ani3D  finite element package~\cite{ani3D}, which was used to generate  tetrahedra subdivisions  of 3D domains, to build stiffness
matrices,  and to implement the linear algebra solvers described in section~\ref{s_solver}.

\section{The 1D-3D coupled model} \label{s_model}
This section reviews 3D and 1D fluid  models and describes the coupling of the models.  In this study, the 3D model is assumed `rigid'.

\subsection{The 3D model} Consider a flow of a viscous incompressible Newtonian fluid in a bounded domain
$\Omega\subset \mathbb{R}^3$. We shall distinguish between the inflow part of the  boundary, $\Gamma_{\rm in}$, the no-slip and no-penetration part (rigid walls), $\Gamma_{0}$,  and the outflow part of the boundary, $\Gamma_{\rm out}$. On the inflow part we assume a given velocity profile. The outflow boundary conditions are defined by setting the normal stress tensor
equal to a given vector function $\mbox{\boldmath$\phi$\unboldmath}$. Thus, the 3D model is the classical  Navier-Stokes equations in pressure-velocity variables:
\begin{equation}  \label{NSE}
\left\{
\begin{split}
\begin{split}
\rho\left(\frac{\partial\bu}{\partial t}+(\bu\cdot\nabla)\,{\bf u}\right)-\nu \Delta {\bf u}  +
\nabla p &={\bf 0}   \\[0.5ex]
\Div {\bf u} &= 0  \end{split}&\quad {\rm in}~\Omega\times(0,T],
\\
\bu|_{\Gamma_{\rm in}} = \bu_{\rm in},\quad\bu|_{\Gamma_{0}} = {\bf 0}&,\\ \left.\left(\nu\frac{\partial\bu}{\partial\bn}-p\bn\right)\right|_{\Gamma_{\rm out}}=\mbox{\boldmath$\phi$\unboldmath}&.
\end{split}
\right.
\end{equation}
Here $\bn$ is the outward normal vector to $\dO$.
The system is also supplemented with initial condition $\bu=\bu_0$ ($\Div\bu_0=0$) for $t=0$ in $\Omega$.

We remark that the notion of `inflow' and `outflow' boundary is used here and further in the text conventionally, since the inequalities $\bu\cdot\bn<0$ or  $\bu\cdot\bn>0$ are \textit{not} necessarily pointwise satisfied  on $\Gamma_{\rm in}$ or  $\Gamma_{\rm out}$, respectively. In applications we  consider, the mean flux, $\bu\cdot\bn$ averaged in space \textit{and in time}, is expected to be negative at $\Gamma_{\rm in}$ and  positive at $\Gamma_{\rm out}$. However, for certain $t\in[0,T]$, the flux $\int\limits_{\Gamma_{\rm in (out)}}\hskip-2ex\bu\cdot\bn\,\rd\bs$
may take positive values at $\Gamma_{\rm in}$ (negative at $\Gamma_{\rm out}$).

If the solution to \eqref{NSE} is sufficiently smooth and the inflow boundary conditions are homogeneous, the following energy
balance holds:
\[
\frac\rho2\frac{\rd}{\rd t}\|\bu\|^2+\nu\|\nabla\bu\|^2
+\int_{\Gamma_{\rm out}}\left(\frac\rho2|\bu|^2\bn-\mbox{\boldmath$\phi$\unboldmath}\right)\cdot\bu \rd\bs=0.
\]
Here and in the rest of the paper, $\|\cdot\|$ denotes the $L^2(\Omega)$ norm. If one \textit{assumes}
\begin{equation}  \label{cond_out}
\int_{\Gamma_{\rm out}}|\bu|^2 \bu\cdot\bn \rd\bs\ge0\quad\forall\,t\in[0,T],
\end{equation}
then solutions to \eqref{NSE} satisfy the \textit{a priori} energy inequality and this opens possibilities for showing partial well-posedness results. Even though, the  assumption \eqref{cond_out}  was used
for the purpose of analysis in the literature on 3D-1D blood flow models (see ~\cite{3D1D1,3D1D2}),
it is hard to verify  \eqref{cond_out} for practical flows. Moreover, the  inequality \eqref{cond_out} no longer holds
if reverse flows occur, as, for example, happens in IVC~\cite{cava1,cava2}.


\subsection{The 1D model}

A one-dimensional model can be derived from the Navier-Stokes equations posed
in a long axisymmetric elastic pipe by integrating over cross section, making some simplifying assumptions and considering integral average quantities as unknowns, see, e.g.,~\cite{Abakumov,QTV}.
Let $\omega(t,x)$ be the  cross section of the pipe normal to $x$, $S(t,x)$ is the area of $\omega(t,x)$
and $u(t,\bx)$ is the axial velocity. Introduce the averaged variables: the mean axial velocity $\bar{u}$ and the mean pressure:
\[
\bar{u}(t,x)=S^{-1}(t,x)\int_{\omega(t,x)}u(t,\bx)\rd\bs,\quad \bar{p}(t,x)=S^{-1}(t,x)\int_{\omega(t,x)}p(t,\bx)\rd\bs.
\]
We consider the model given by the following system of equations for unknowns $\bar{u},\bar{p},S$:
\begin{equation}\label{1Dmodel}
\left \{\begin{split}
\frac{\partial S}{\partial t}+\displaystyle\frac{\partial (Su)}{\partial x}&=\varphi(t,x,S,\bar{u})\\
\frac{\partial \bar{u}}{\partial
t}+\displaystyle\frac{\partial (\bar{u}^2/2+\bar{p}/\rho)}{\partial
x}&=\psi(t,x,S,\bar{u})\\
  \bar{p}-p_{\rm ext}&=\rho c^2_0f(S)
  \end{split}\right.\qquad\text{for}~~x\in[0,1].
\end{equation}
For  initial conditions, the mean velocity and the cross section area are prescribed,
$\bar{u}|_{t=0}=u_0$, $S|_{t=0}=S_0$. Here $p_{\rm ext}$ is the external pressure,  $\varphi(t,x,S,\bar{u})$ is a function modelling the source or sink of the fluid, as may be required in  hemodynamic simulations, if a blood loss happens in a vessel. Further, we assume
$\varphi=0$ and $p_{\rm ext}=0$, so from now $\bar{p}$ has the meaning of the difference between the fluid pressure and the external pressure. The term $\psi(t,x,S,\bar{u})$ accounts for external forces, such as gravity or friction. Following~~\cite{Abakumov,Simakov}, we set
\begin{equation}\label{func_psi}
\psi=-16\nu \bar{u}\eta(\widetilde{S})(\widetilde{S} d^2)^{-1},\quad \widetilde{S}=\widehat{S}^{-1}S.
\end{equation}
Here $\nu$ is the viscosity coefficient, $d$ is the pipe  diameter, $\widehat S$ is the reference area (in the hemodynamic applications $\widehat S$ is the cross section area of a vessel at rest) and
\[
\eta(\widetilde{S})=\left \{
\begin{split}
2, &\quad\text{for}~ \widetilde{S}>1\\
\widetilde{S}+ \widetilde{S}^{-1},&\quad\text{for}~ \widetilde{S}\le1.
\end{split}\right.
\]
The last equation in \eqref{1Dmodel} relates the pressure to the cross section area. The function $f$
is defined by the elasticity model of the pipe walls, $c_0$ is the elasticity model parameter. We use the one from \cite{Simakov}:
\begin{equation}\label{func_f}
f(S)=\left \{
\begin{split}
\exp{(S\widehat{S}^{-1}-1)}-1, &\quad\text{for}~ S>\widehat{S}\\
\ln{(S\widehat {S}^{-1})},&\quad\text{for}~ S\le\widehat{S}.
\end{split}\right.
\end{equation}
Other algebraic defining relations linking  the mean pressure and the cross section area are known
from the literature, see, e.g., \cite{QTV}. They are equally well suited for the purpose of this paper.

In \cite{3D1D1}, the authors derived the energy equality for the one-dimensional fluid
model in  ${Q,S,\bar{p}}$ variables, where $Q=S\bar{u}$ is the flux. Up to possibly different
choices of $\psi(t,x,S,\bar{u})$ and $f(S)$, the formulation in \cite{3D1D1} is  equivalent to \eqref{1Dmodel}. Written in
terms of $\bar{u},S$, and $\bar{p}$, the energy equality for \eqref{1Dmodel} is  (recall that
we assumed $\varphi=0$):
\begin{equation}\label{energy1D}
\frac{\rm d}{\rm dt}\cE_{1D}(t) - \rho\int_0^1 S \psi(t,x,S,\bar{u})\, \bar{u}{\rm dt}=
-\left.S\bar{u}(\bar{p}+\frac\rho2 \bar{u}^2)\right|_{0}^1,
\end{equation}
with the energy functional
\[
\cE_{1D}(t)=\frac\rho2 \int_0^1S\bar{u}^2{\rm dx}+ \rho c^2\int_0^1 \int_{\widehat{S}}^Sf(s){\rm d}s {\rm dx}.
\]
For $f(S)$ given in \eqref{func_f}, the second term in the definition of $\cE_{1D}(t)$ is always positive, making $\cE_{1D}(t)$ positive for all $t>0$. The choice of $\psi(t,x,S,\bar{u})$ in \eqref{func_psi}
ensures that the second term  on the left-hand side of \eqref{energy1D} is positive as well. Thus, for the homogenous boundary conditions the
energy of the 1D model dissipates: $\frac{\rm d}{\rm dt}\cE_{1D}(t)<0$.
\medskip

System \eqref{1Dmodel} is hyperbolic and can be integrated along characteristics.
To see this, we write \eqref{1Dmodel} in the divergence form:
\[
\frac{\partial V}{\partial t}+\frac{\partial F(V)}{\partial
x}=g,
\]
with  $V=\{S,\bar{u}\}$, $\displaystyle F=\{S\bar{u}, \frac{\bar{u}^2}2+\frac{\bar{p}}{\rho}\}$,
$g=\{0,\psi\}$. Denote by $\mathbf{w}_{i}$ and  $\lambda_{i}$ (i=1,2) the left eigenvectors and eigenvalues of the Jacobian $A=\frac{\partial F}{\partial V}$. One finds
$$
\lambda_{i}=\left\{
\begin{split}
\bar{u}
+(-1)^i c_0\sqrt{S/{\widehat S}\exp({S}/{\widehat S}-1)}&\quad\text{if}~ S>\widehat{S}\\
\bar{u}
+(-1)^i c_0&\quad\text{if}~S\le\widehat{S}
\end{split}
\right.,\quad  i=1,2,
$$
and
\[\mathbf{w}_{i} = \left \{
\begin{split}
 \left(c_0\sqrt{(1/\widehat S)\exp(S/\widehat S - 1)},\,(-1)^i\sqrt S\right)^T
&\quad\text{if}~ S>\widehat{S}\\
\left(c_0,\,(-1)^i S\right)^T &\quad\text{if}~S\le\widehat{S}
\end{split}
\right.,\quad i=1,2.
\]
Thus, system \eqref{1Dmodel} can be written in the characteristic form:
\begin{equation}
\label{111}
\mathbf{w}_{i}\left(\frac{\partial V}{\partial
t}+\lambda_{i}\frac{\partial V}{\partial x}\right)=\mathbf{w}_{i}g,\quad
i=1,2.
\end{equation}
Under physiological conditions in hemodynamics, it holds $\bar{u}<c_0$, implying that the eigenvalues $\lambda_i$ have
opposite signs. Therefore, $\mathbf{w}_{1}$ is the incoming characteristic from point $x=1$ and
$\mathbf{w}_{2}$ is the incoming characteristic from point $x=0$. Two boundary conditions, one at
$x=0$ and one at $x=1$, are enough to close the system.

\subsection{The coupling of 1D and 3D models}\label{s_coupling}

\begin{figure}\centering
\begin{picture}(300,130)(0,0)
\put(0,0){\includegraphics[width=0.75\textwidth, trim = 0 150 0 150, clip]{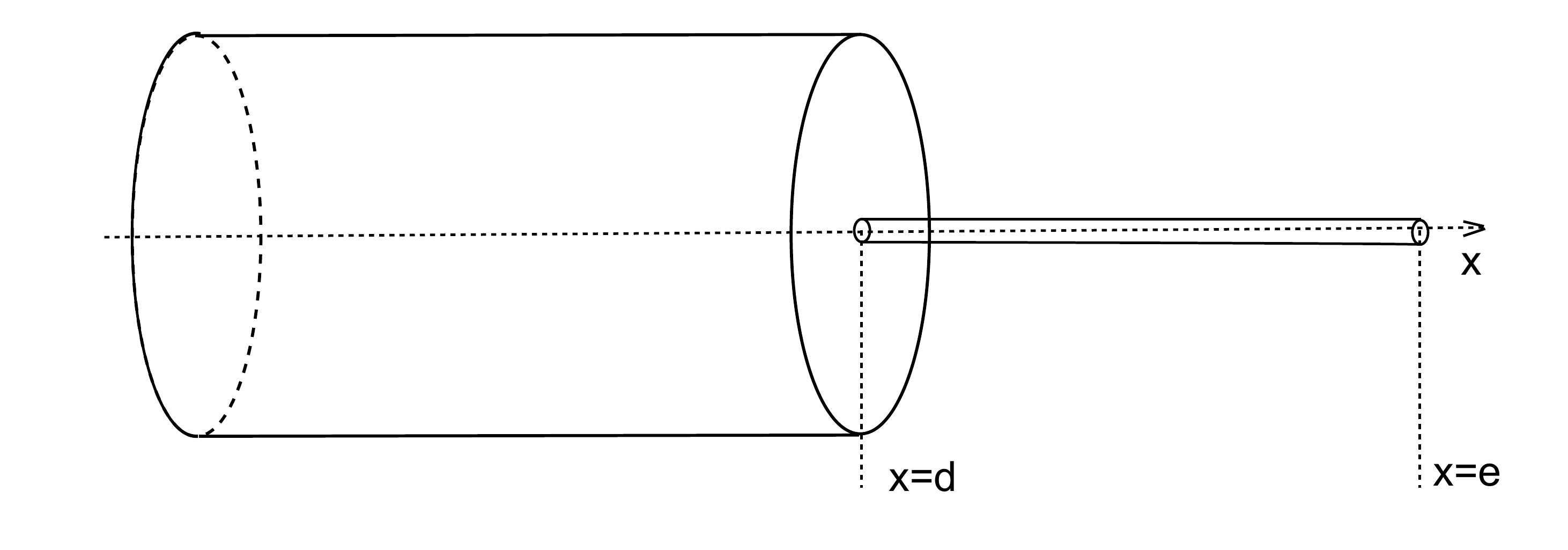}}
\put(100,50){$\Omega_{\rm 3D}$}
\put(230,50){$\Omega_{\rm 1D}^{\rm down}$}
\put(166,78){$\Gamma_{\rm out}$}
\put(36,78){$\Gamma_{\rm in}$}
\end{picture}
\caption{ The schematic coupling   of  $\Omega_{\rm 3D}$ and $\Omega_{\rm 1D}^{\rm down}$
domains.} \label{fig1}
\end{figure}
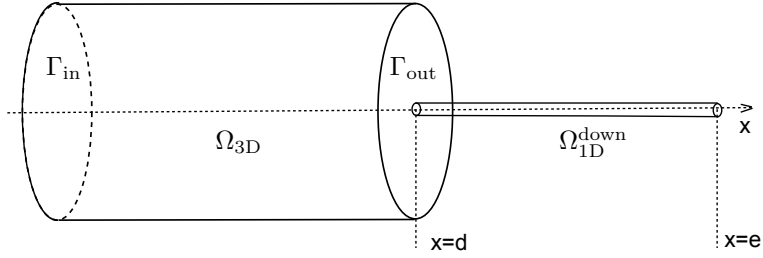

We now consider two domains  $\Omega_{\rm 3D}$ and $\Omega_{\rm 1D}^{\rm down}$, as shown in
Figure~\ref{fig1}.
 In  $\Omega_{\rm 1D}^{\rm down}$ we pose the
1D fluid model and in $\Omega_{\rm 3D}$ we pose the full three-dimensional Navier-Stokes equations.
The one-dimensional domain $\Omega_{\rm 1D}^{\rm down}$ is coupled to the downstream boundary of $\Omega_{\rm 3D}$.

There are several  options to define the coupling conditions of 1D and 3D models, as discussed, for example, in \cite{3D1D1}.  One can ask for the continuity of the mean pressure,
the mean axial velocity, the flux, the normal cross section  area
, the averaged normal stress, or the entering characteristic.
In general, the continuity of all these quantities cannot be satisfied simultaneously. A choice has to be made.

One common choice, see, e.g., \cite{UBVF,BFU},  is to impose the continuity of the normal stress and the  flux:
\begin{align}\label{cond_b1}
\left.\Big(-\nu\frac{\partial {\bf u} }{ \partial {\bn}} + {
p}\bn\Big)\right|_{{\Gamma}_{\rm out}}& = \bar{p}|_{x=d}\bn,\\
 \int_{{\Gamma}_{\rm out}} {\bf u}
\cdot {\bf n}\, ds& = S\bar{u}|_{x=d}. \label{cond_b2}
\end{align}
This choice is, however, known to be energy inconsistent in the following sense. Assume the homogeneous boundary conditions on $\Gamma_{\rm in}$ and the downstream end of $\Omega_{\rm 1D}^{\rm down}$. Then
the cumulative  energy balance of the coupled model is
\begin{multline}\label{energy}
\frac{\rm d}{\rm dt}(\cE_{3D}(t)+\cE_{1D}(t)) +\nu\|\nabla\bu\|^2+\int_{\Omega_{\rm 1D}^{\rm down}}K_\nu(S) \bar{u}^2{\rm dx}=\\
\int\limits_{\Gamma_{out}} \left(\nu\frac{\partial\bu}{\partial\bn} - (p+\frac\rho2|\bu|^2)\bn\right)\cdot\bu\rd\bs+\left.S\bar{u}(\bar{p}+\frac\rho2 \bar{u}^2)\right|_{x=d},
\end{multline}
with $\cE_{3D}(t)=\frac12\|\bu\|^2$. $K_\nu(S)$ is a positive coefficient defined from \eqref{func_psi}. Easy to see that for coupling conditions \eqref{cond_b1}--\eqref{cond_b2} the right hand side of \eqref{energy}
reduces to
\[
\frac\rho2\left(\left.Su^3\right|_{x=d}-\int\limits_{\Gamma_{out}}|\bu|^2(\bu\cdot\bn)\rd\bs\right).
\]
In general, it is not clear if this quantity is non-positive and thus if the cumulative  energy is properly dissipated, as it holds for the full 3D Navier-Stokes equations with the Dirichlet homogeneous boundary conditions. To circumvent this inconsistency, it was suggested in \cite{3D1D2} to replace  condition \eqref{cond_b1} by the continuity of so-called total stress:
\begin{equation}\label{totalstresscoupling}
-\nu\frac{\partial\bu}{\partial\bn} + (p+\frac\rho2|\bu|^2)\bn = \left.(\bar{p}+\frac\rho2 \bar{u}^2)\right|_{x=d} \bn \quad \mbox{ on } \Gamma_{\rm out}.
\end{equation}
Condition  \eqref{totalstresscoupling} together with the continuity of the flux, i.e.  condition  \eqref{cond_b2}, makes the right hand side of \eqref{energy} equal to zero. Hence, the cumulative energy dissipates. However, setting the total stress equal to a constant is not a consistent outflow condition for the simplest Poiseuille flow. Moreover,  \eqref{totalstresscoupling} is the natural boundary condition for the \textit{rotation} form of the Navier-Stokes equations. Using it with the common convection or conservation forms leads to non-linear coupling conditions
and often requires iterative treatment~\cite{BSI}.   Although the rotation form is an interesting alternative to the standard convection form,  it is still not a standard option in the existing software and its use requires certain  care \cite{LMNOR,O2000}.

The condition for the normal stress, as in \eqref{cond_b1}, is the natural boundary condition for the commonly used  convection and conservation forms of the Navier-Stokes equations. Such a condition has been shown to be surprisingly  useful as outflow boundary condition~\cite{HR97}. Thus, instead of  keeping \eqref{cond_b2} and changing   \eqref{cond_b1} to the total stress condition, we retain \eqref{cond_b1} and instead of   \eqref{cond_b2} assume the continuity of the linear combination of the fluid flux and the energy flux:
 \begin{equation}\label{newfluxcoupling}
 \bar{p}(d)\int\limits_{\Gamma_{out}} \bu\cdot \bn\rd\bs +\frac\rho2\int\limits_{\Gamma_{out}}|\bu|^2(\bu\cdot\bn)\rd\bs =  (\bar{p}S\bar{u} +\frac \rho2 S\bar{u}^3)|_{x=d}.
\end{equation}
One easily checks that the combination of  \eqref{cond_b1} and   \eqref{newfluxcoupling} ensures the right-hand side of \eqref{energy} equal to zero and thus
the correct energy balance and energy inequality are valid as formulated in the following theorem.

\begin{theorem}\label{Th1}
Consider the coupled 1D--3D fluid problem given by \eqref{NSE}, \eqref{1Dmodel}, \eqref{func_psi}, \eqref{func_f},
and coupling conditions  \eqref{cond_b1}, \eqref{newfluxcoupling}. Assume the homogeneous boundary conditions $\bu_{\rm in}=0$ and $\bar{u}|_{x=e}=0$. Then a sufficiently smooth solution satisfies the following energy decay property:
 \begin{equation}\label{energy_balance}
\cE_{3D}(t)+\cE_{1D}(t) +\nu\int_0^t\|\nabla\bu\|^2{\rm dt'}+\int_0^t\int_{\Omega_{\rm 1D}^{\rm down}} K_\nu(S)\bar{u}^2{\rm dx}{\rm dt'}=
\cE_{3D}(0)+\cE_{1D}(0)
\end{equation}
for any $t\in[0,T]$. If $\psi$ is defined in \eqref{func_psi}, then $K_\nu(S)=16\nu\eta(\widetilde{S})\widehat{S}d^{-2}>0$.
\end{theorem}

\begin{remark} \rm
Since $\bar{p}$ has the meaning of the difference between fluid and external pressures, it can be negative. In this case,  more than one value of  $\bar{u}$ may satisfy \eqref{newfluxcoupling}.
To ensure that the coupled model is not defective,
one has to prescribe a particular rule for choosing the root of the cubic equation \eqref{newfluxcoupling}.
In our numerical experiments, we take  $\bar{u}$ which is the closest to $|\Gamma_{out}|^{-1}\int\limits_{\Gamma_{out}} \bu\cdot \bn\rd\bs$.
\end{remark}

\begin{remark} \rm
The boundary condition  \eqref{newfluxcoupling} is the combination of fluid and energy fluxes and so it does not
guarantee to conserve the `mass' of the entire coupled system. Although, we do not observe any   perceptible generation or loss of mass in our numerical experiments, it does not necessarily mean that for all problems this effect should be negligible. Actually, one may consider any other linear combination of fluid and energy fluxes coupling on 3D-1D boundary to compromise between energy stability and mass conservation.
\end{remark}

\begin{figure}\centering
\begin{picture}(300,130)(0,0)
\put(0,0){\includegraphics[width=0.75\textwidth, trim = 0 150 0 150, clip]{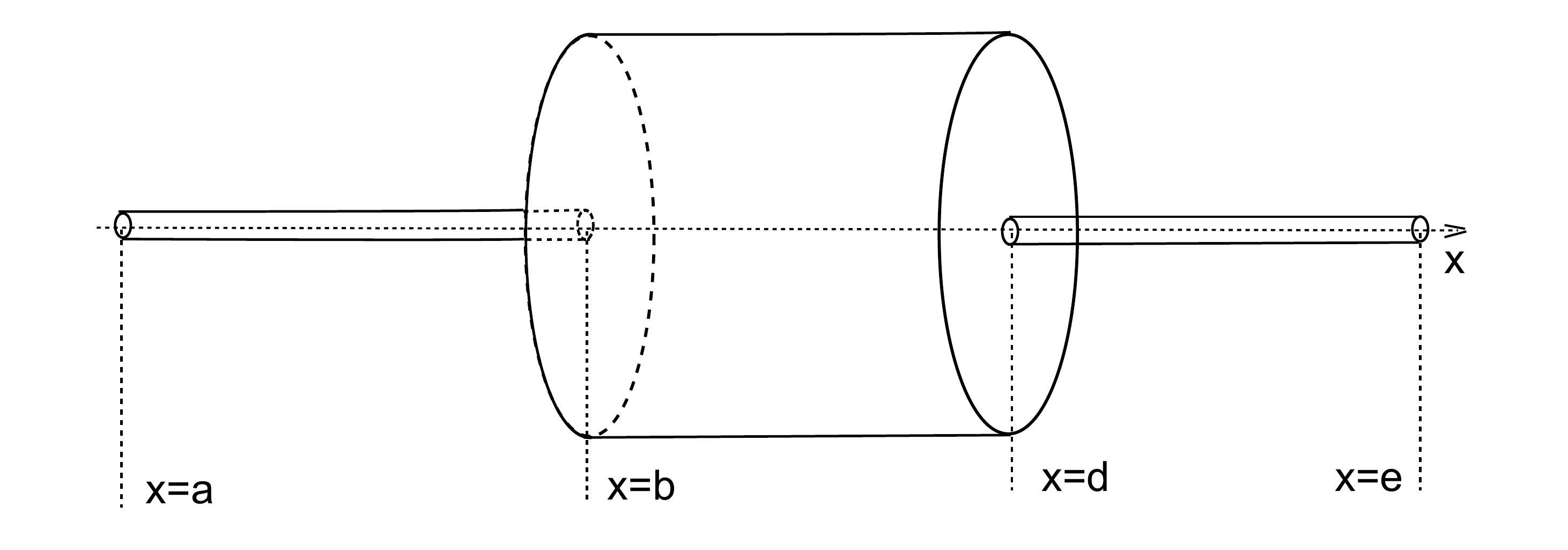}}
\put(60,50){$\Omega_{\rm 1D}^{\rm up}$}
\put(155,50){$\Omega_{\rm 3D}$}
\put(240,50){$\Omega_{\rm 1D}^{\rm down}$}
\put(198,78){$\Gamma_{\rm out}$}
\put(115,78){$\Gamma_{\rm in}$}
\end{picture}
\caption{ The schematic coupling   of  $\Omega_{\rm 1D}^{\rm up}$, $\Omega_{\rm 3D}$ and $\Omega_{\rm 1D}^{\rm down}$
domains.} \label{fig2}
\end{figure}
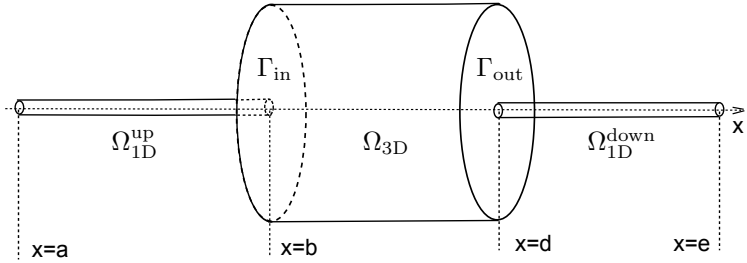

In practice, one may also be interested in coupling the 1D fluid model to the \textit{upstream} boundary of the 3D domain.
Hence, we now consider three domains $\Omega_{\rm 1D}^{\rm up}$, $\Omega_{\rm 3D}$, and $\Omega_{\rm 1D}^{\rm down}$, as shown in Figure~\ref{fig2}. In $\Omega_{\rm 1D}^{\rm up}$ and $\Omega_{\rm 1D}^{\rm down}$  the simplified
1D model is posed and in $\Omega_{\rm 3D}$  the full three-dimensional Navier-Stokes equations are solved.
The  domain $\Omega_{\rm 1D}^{\rm up}$ is coupled to the inflow (upstream) boundary of $\Omega_{\rm 3D}$ and $\Omega_{\rm 1D}^{\rm down}$ is coupled to the outflow (downstream) boundary of $\Omega_{\rm 3D}$. The downstream coupling is described above. In the literature, it is common not to distinguish between upstream and downstream
coupling boundary conditions. For example, in \cite{BFU,UBVF} conditions \eqref{cond_b1},\eqref{cond_b2} are
assumed both on upstream and downstream boundaries. Following this paradigm, one may consider  \eqref{cond_b1},\eqref{cond_b2} or energy consistent conditions \eqref{cond_b1}, \eqref{newfluxcoupling} as the coupling
conditions on $\Gamma_{\rm in}$ between   1D model posed in $\Omega_{\rm 1D}^{\rm up}$ and 3D model posed in $\Omega_{\rm 3D}$. Note that in  entirely 3D fluid flows simulations, inflow and outflow boundary conditions usually differ.
If a numerical approach to 1D-3D problem is based on  subdomains splitting (see the next section for an example), then it is
 appropriate to distinguish between upstream and downstream coupling conditions. Thus, we impose the upstream coupling conditions in such a way that the 3D problem is supplied with the Dirichlet inflow boundary conditions. This is a standard choice for incompressible viscous fluid flows  solvers and
is especially convenient  if third parties or legacy codes are separately used to compute 3D and 1D solutions,
and they communicate only through coupling conditions.

For the upstream boundary, ${\Gamma}_{\rm in}$ , we  introduce a reference velocity profile $\tilde{\bg}(\bx)$, $\bx\in {\Gamma}_{\rm in}$,  such that $\int_{{\Gamma}_{\rm in}} \tilde{\bg} \cdot {\bn}\, \rd\bs=1$.
Then the boundary condition on  ${\Gamma}_{\rm in}$ is Dirichlet, given by
\begin{equation}\label{cond_a1}
\bu_{\rm in}=-\alpha\tilde{\bg}\quad\text{on}~{\Gamma}_{\rm in}.
\end{equation}
Setting $\alpha=S\bar{u}|_{x=b}$ ensures the continuity of the flux \eqref{cond_b2} on $\Gamma_{in}$. If $\alpha$ is found to satisfy the equation
\begin{equation*}
 \bar{p}(b)\alpha +\frac\rho2\alpha^3\int\limits_{\Gamma_{in}}|\tilde{\bg}|^2(\tilde{\bg}\cdot\bn)\rd\bs =  (\bar{p}S\bar{u} +\frac \rho2 S\bar{u}^3)|_{x=b},
\end{equation*}
then the coupling condition \eqref{newfluxcoupling} is valid on $\Gamma_{in}$.
Two more  scalar boundary conditions are required for the 1D model in $\Omega_{\rm 1D}^{\rm up}$.
We assume that $\bar{u}$ or $\bar{p}$ are given in $x=a$ and in $x=b$
an absorbing condition is prescribed: in computations we set $(Su)_x=0$ in $x=b$; another reasonable absorbing condition would be setting the incoming characteristic equals zero.
On the downstream end of $\Omega_{\rm 1D}^{\rm down}$, we also assume an absorbing boundary condition.

We summarize the properties for the 3D-1D coupling introduced in this section:
\begin{itemize}
\item It ensures the energy balance, as stated in Theorem~\ref{Th1};
\item The inequality \eqref{cond_out} is not assumed;
\item It can be easy decoupled with splitting methods into the separate 1D problems and the 3D problem with usual inflow-outflow
boundary conditions on every time step.
\end{itemize}

\section{Discretization and algebraic solver} ~\label{s_solver}

In this section, we introduce a splitting numerical time-integration algorithm based on subdomain splitting. Further, we consider a fully discrete
problem and review one state-of-the-art algebraic solver.

Denote by $\bar{u}^n,\bar{p}^n$, $S^n$, $\bu^{n}$, and $p^n$ approximations to the corresponding
unknown variables at time $t=t_n$.
 Given these  approximations, we compute $\bar{u}^{n+1},\bar{p}^{n+1}$, $S^{n+1}$, $\bu^{n+1}$, and $p^{n+1}$ for $t=t_{n+1}$ ($\Delta t=t_{n+1}-t_{n}$) in three steps:

\begin{description}
\item{Step~1.} Integrate (\ref{111}) for $t\in[t_n,t_{n+1}]$, with given $\bar{u}(t_{n+1})$ on the upstream end of the interval $\Omega_{\rm 1D}^{\rm up}$  and the absorbing downstream condition at $x=b$.

\item{Step~2.} Set $\bu_{\rm in}$ according to \eqref{cond_a1}, with $\{\bar{u},\bar{p},S\}=\{\bar{u}^{n+1},\bar{p}^{n+1},S^{n+1}\}$, and  compute $\bar{p}^\ast$ and $S^\ast$ as the linear extrapolations of $\bar{p}|_{x=d}$ and $S|_{x=d}$ from times $t_n$ and $t_{n-1}$. Solve the linearized
    Navier-Stokes problem in  $\Omega_{\rm 3D}$ for $\bu^{n+1},{p}^{n+1}$:
\begin{equation*}  
\left\{
\begin{split}
\frac1{2\triangle t}( 3{\bu^{n+1}}- 4\bu^{n}+\bu^{n-1}) + (2\bu^{n}-\bu^{n-1})\cdot \nabla  {\bu^{n+1}} - \nu\Delta  {\bu^{n+1}} + \nabla p^{n+1}& = \blf^{n+1}, \\
\Div {\bf u}^{n+1} &= 0, \\
 {\bu^{n+1}}|_{\Gamma_{\rm in}}= \bu_{\rm in},\quad {\bu^{n+1}}|_{\Gamma_{0}}= 0,\quad \Big(-\nu\frac{\partial\bu^{n+1}}{\partial\bn}+p^{n+1}\, \bn\Big)|_{\Gamma_{\rm out}}&=\bar{p}^\ast\bn.
\end{split}
\right.
\end{equation*}
\item{Step~3.} Find $\bar{u}^{n+1}|_{x=d}$ from
\[
(\bar{p}^\ast S^\ast \bar{u}^{n+1} +\frac \rho2 S^\ast (\bar{u}^{n+1})^3)|_{x=d}=
 \bar{p}(d)\int\limits_{\Gamma_{out}} \bu^{n+1}\cdot \bn\rd\bs +\frac\rho2\int\limits_{\Gamma_{out}}|\bu^{n+1}|^2(\bu^{n+1}\cdot\bn)\rd\bs .
\]
Now, using $\bar{u}^{n+1}$ for boundary condition in $x=d$ and the absorbing boundary condition in $x=e$, we integrate (\ref{111}) for $t\in[t_n,t_{n+1}]$ to find $\bar{u}^{n+1},\bar{p}^{n+1}$, $S^{n+1}$ in $\Omega_{\rm 1D}^{\rm down}$.
\end{description}

For the numerical integration of  the 1D model equations, we use a first order monotone
finite difference  scheme applied to  the characteristic form \eqref{111}, see~\cite{all1}.
To handle the 3D model, one has to solve on every time step  the linearized Navier-Stokes equations,
also known as the Oseen problem:
\begin{equation}  \label{Oseen}
\left\{
\begin{split}
\begin{split}
\beta\bu-\nu \Delta {\bf u} + (\bw\cdot\nabla)\,{\bf u} +
\nabla p &={\bf f}   \\[0.5ex]
\Div {\bf u} &= 0  \end{split}&\qquad {\rm in}~\Omega_{\rm 3D},
\\
{\bf u}|_{\Gamma_{\rm in}\cup\Gamma_{0}} = {\bf g},\quad (\nu\frac{\partial\bu}{\partial\bn}-p\bn)|_{\Gamma_{\rm out}}=0 &
\end{split}
\right.
\end{equation}
where $\bu=\bu^{n+1}$, $p=p^{n+1}-\bar{p}^\ast(d)$, the body forces term and the advection velocity field depend on previous time velocity approximations, ${\bf f}=(2\Delta t)^{-1}(4\bu^{n}-\bu^{n-1})$, $\bw=(2\bu^{n}-\bu^{n-1})$, and
 $\beta=3(2\Delta t)^{-1}$. Here and in the remainder of this section, we dropped out the
time-step dependence ($n+1$) index for unknown velocity and pressure.

To discretize the Oseen problem
\eqref{Oseen}, we consider a conforming finite element method.
Denote the finite element velocity  and pressure spaces by
$\VV_h\subset H^1(\Omega_{\rm 3D})^3$ and $\QQ_h\subset L^2(\Omega_{\rm 3D})$, respectively.
Let $\VV_h^0$ be the subspace of $\VV_h$ of all FE velocity functions vanishing at $\Gamma_{\rm in}\cap\Gamma_0$.

The finite element problem reads: Find $\bu_h\in\VV_h$, $\bu_h|_{\Gamma_{\rm in}}=\bu_{\rm in}^h$,
and $p_h\in\QQ_h$ satisfying
\begin{equation} \label{FE_Oseen}
a(\bu_h,\bv_h)-(p_h,\Div\bv_h)+(q_h,\Div\bu_h)
=(\blf_h,\bv_h)\quad\forall~\bv_h
\in\VV_h^0,\,q_h\in\QQ_h,
\end{equation}
with
\[
a(\bu_h,\bv_h)=\beta(\bu_h,\bv_h) +\nu(\nabla\bu_h,\nabla\bv_h)+(\bw\cdot\nabla\bu_h,\bv_h).
\]
Let $(\psi,\phi)_\VV=(\nabla\psi,\nabla\phi)$ for $\psi,\phi\in\VV_h^0$. We assume the ellipticity, the continuity, and
the stability conditions:
\begin{equation}
\beta_1\|\bv_h\|^2_\VV\le a_h(\bv_h,\bv_h),\qquad
a(\bv_h,\bu_h)\le \beta_2\|\bv_h\|_\VV\|\bu_h\|_\VV \quad\forall\
\bv_h,\bu_h\in\VV_h^0, \label{el_cont}
\end{equation}
\begin{equation}
\gamma_1^2\,\|q_h\|^2\le\sup_{\bv_h\in\VV_h^0}\frac{(q_h,\Div\bv_h)^2}{\|\bv_h\|_\VV^2}\quad\forall\
q_h\in\QQ_h,
\label{LBBm}\end{equation}
\begin{equation}
(q_h,\Div\bv_h)\le\gamma_2\|q_h\|\|\bv_h\|_\VV \quad\forall\
q_h,p_h\in\QQ_h,~\bv_h\in\VV_h^0, \label{cont_c}
\end{equation}
with positive mesh-independent constants $\beta_1$, $\beta_2$,
$\gamma_1$, and $\gamma_2$. Condition
\eqref{LBBm} is well-known as the LBB or inf-sup stability condition~\cite{BF}.

Let $\{\phi_i\}_{1\le i\le n}$ and $\{\psi_j\}_{1\le j\le m}$ be
bases of $\VV_h^0$ and $\QQ_h$, respectively. Define the following
matrices:
\[
A_{i,j}=a(\phi_j,\phi_i),\quad B_{i,j}=-(\Div\phi_j,\psi_i).
\]
The linear algebraic system corresponding to \eqref{FE_Oseen}
(\textit{the discrete Oseen system}) takes the form
\begin{equation} \label{sp}
\left (\begin{array}{cc} A & B^T \\ B & 0 \end{array}\right ) \left
(\begin{array}{c} u \\ p \end{array}\right ) = \left
(\begin{array}{cc} f \\ g \end{array}\right )\,.
\end{equation}
The right hand side $(f,g)^T$ accounts for body forces and inhomogeneous velocity boundary conditions.
To solve \eqref{sp}, we consider a Krylov subspace iterative method, with the
block triangular preconditioner~\cite{ElmSilWat05BOOK,KS}:
\begin{equation} \label{btal} {\cal P}=
\left (\begin{array}{cc} \widehat A & B^T \\
O & -\widehat S\end{array}\right ) \,.
\end{equation}
The matrix ${\widehat A}$ is a preconditioner for the matrix $A$, such
that ${\widehat A}^{-1}$ may be considered as an inexact solver for
linear systems involving $A$. The matrix $\widehat S$ is a
preconditioner for the pressure Schur complement of (\ref{sp}),
$S=BA^{-1}B^T$. In the algorithm, one needs the actions
of ${\widehat A}^{-1}$ and ${\widehat S}^{-1}$ on subvectors, rather than
the matrices ${\widehat A}$, $\widehat S$ explicitly. Once good
preconditioners for $A$ and $S$ are available, a preconditioned Krylov subspace method, such as GMRES
or BiCGstab, is the efficient solver. In the
literature, one can find geometric or algebraic multigrid (see, e.g.,
\cite{ElmSilWat05BOOK} and references therein)
or domain decomposition \cite{Garbey,Vally}  algorithms
which provide effective preconditioners $\widehat A$ for a  range of
$\nu$ and various meshes.
We use one V-cycle of the algebraic multigrid method~\cite{Stuben} to define ${\widehat A}^{-1}$.

Defining an appropriate pressure Schur complement preconditioner ${\widehat S}^{-1}$ is more challenging.
In this paper, we follow the approach of Kay et al.~
\cite{KayLogWat02}.
First, we define the pressure mass and velocity mass matrices:
\[
(M_p)_{i,j}=(\psi_j,\psi_i),\quad (M_u)_{i,j}=(\phi_j,\phi_i).
\]
The original pressure convection-diffusion (PCD) preconditioner,
proposed in \cite{KayLogWat02}, is defined through its inverse:
\begin{equation}\label{PCD}
\widehat S^{-1} := \widehat{M}_p^{-1}A_pL_p^{-1}.
\end{equation}
Here $\widehat M_p^{-1}$ denotes an approximate solve with the pressure
mass matrix. Matrices $A_p$ and $L_p$ are approximations to
convection-diffusion and Laplacian operators in $\QQ_h$,
respectively. Both $A_p$ and $L_p$ (explicitly or implicitly) assume
some pressure boundary conditions to be prescribed.

If $\QQ_h$ defines continuous pressure approximations, one can use the conforming discretization of
the pressure Poisson problem with  Neumann boundary conditions:
\[
(L_p)_{i,j}=(\nabla \psi_j, \nabla \psi_i).
\]
Likewise, Neumann boundary conditions are conventionally used to define  the pressure convection-diffusion problem on $\QQ_h$. However,  the optimal boundary conditions setup both
for $L_p$ and $A_p$ depends on the type of the boundary
and flow regime, see \cite{Tuminaro,OV07}.

We use a modified PCD preconditioner defined below. This modification partially obviates the
issue of setting pressure boundary conditions and is consistent with the Cahouet--Chabard
preconditioner~\cite{CC}, if the inertia terms are neglected. The  Cahouet--Chabard
preconditioner is the standard choice for the time-dependent Stokes problem and
enjoys the solid mathematical analysis in this case~\cite{OR06}.
To define the preconditioner,  we introduce the discrete advection matrix for continuous pressure
approximations as
 \[
(N_p)_{i,j}=(\bw\cdot\nabla \psi_j, \psi_i).
\]
Then the modified pressure convection-diffusion preconditioner (mPCD)
is (compare to \eqref{PCD}\,):
 \begin{equation*}
\widehat S^{-1} := \nu\widehat{M}_p^{-1}+(\beta I+N_p) (B\widehat{M}_u^{-1}B^T)^{-1},
\end{equation*}
where $\widehat{M}_u$ is a diagonal approximation to the velocity mass matrix.
%

Regarding the numerical analysis of the algebraic solver used here, we note
the following. The eigenvalues bounds of the preconditioned Schur complement:
\begin{equation}
0<c_1 \le|\lambda(S \widehat{S}^{-1})|\, \le C_1, \label{est1}
\end{equation}
were proved for $\beta=0$ and the LBB stable finite elements in \cite{Loghin} and for a more general case in~\cite{OV07}.
The constants $c_1,  C_1$ are independent of the
meshsize $h$, but may depend on the ellipticity, continuity and stability constants in \eqref{el_cont}--\eqref{cont_c}, and thus  may depend on the problem parameters.
In particular, the pressure stability constant $\gamma_1$, and so $c_1$ from \eqref{est1},
depends on the geometry of the domain $\Omega$~\cite{ChOl} (tending to zero for long or narrow domains)
and for certain FE pairs $\gamma_1$ depends on the anisotropy ratio of a triangulation~\cite{Apel}.
Both of this dependencies require certain care in using the approach for computing flows in  3D elongated domains with thin and anisotropic inclusions (prototypical for simulating a flow over  IVC filter).

Characterizing the rate of convergence of
nonsymmetric preconditioned iterations is a difficult task. In
particular, eigenvalue information alone may not be sufficient to
give meaningful estimates of the convergence rate of a method like
preconditioned GMRES \cite{GPS}. Nevertheless, experience shows that for
many linear systems arising in practice, a well-clustered spectrum
(away from zero) usually results in rapid convergence of the
preconditioned iteration. This said, we should mention that a
rigorous proof of the GMRES convergence applied to \eqref{sp}, with
block-triangular preconditioner \eqref{btal}, is not available in the literature
(except the special case, when $\widehat{S}$ is symmetric~\cite{KS,BO11}). Thus, the numerical
assessment of the approach is of  practical interest.

\section{Accuracy and efficiency assessment} \label{s_validate}

In this section, we validate the accuracy and stability of the solver for  the 3D-1D coupled fluid model
by (i) comparing the computed discrete solutions against an analytical solution for a problem with
simple geometry; (ii) computing the drag coefficient and the pressure drop value for  the flow around the
3D circular cylinder.
The Taylor-Hood P2-P1 elements were used for the velocity-pressure approximation.
The resulting linear algebraic  systems \eqref{sp} were solved by the preconditioned BiCGstab method.
The initial guess in the BiCGstab method  was zero on the first time step and equal to
the $(\bu^n,p^n)$ for the subsequent time steps. The stopping criteria was
the $10^{-6}$ decrease of the Euclidean norm of the residual.

\subsection{Test with analytical solution}
First we consider an example with analytical solution. The 3D domain is  $\Omega_{\rm 3D}=\{\bx\in\mathbb{R}^3\,|\,x\in(-1,1),\,y^2+z^2<1\}$. The circular cross sections are the
inflow and outflow boundaries.
Domains $\Omega_{\rm 1D}^{\rm up}$ and $\Omega_{\rm 1D}^{\rm down}$ are two intervals of length 5.
The analytical solution is given by
\begin{equation}\label{solution}
\left\{
\begin{split}
S=
\cos {(2{\pi}t)} + \widehat{S}-1,\quad u=1-\cos{(2{\pi}t)},\quad \bar{p}= c^2f(S),\quad\mbox{in}~\Omega_{\rm 1D}^{\rm up}\cup\Omega_{\rm 1D}^{\rm down},\\
\bu=\left( \frac{2S}{\pi}(1-\cos{(2{\pi}t)})(1-y^2-z^2),0,0\right)^T,\quad p=10(1-x)+\bar{p},\quad\mbox{in}~
\Omega_{\rm 3D},
\end{split}
\right.
\end{equation}
with $\widehat{S}=\pi$, $\rho=1$, $c=350$, $\nu=1$. This solution satisfies the continuity of flux condition on the
coupling boundaries. 
The right-hand sides $\varphi$, $\psi$ and $\blf$ were set accordingly.
In this test, the 3D domain was triangulated using the global refinement of an initial mesh, resulting in the sequence of meshes (further denoted by mesh~1, mesh~2, mesh~3), with the number of tetrahedra $N_{tet}=1272, 8403, 63384$, respectively. Since we use the first order scheme for the 1D
problem, the mesh size in $\Omega_{\rm 1D}^{\rm up}$ and $\Omega_{\rm 1D}^{\rm down}$ was divided by 4 on each level of refinement: $\Delta x=5/16, 5/64, 5/256$.
 The corresponding time step was halved for every spacial refinement, so we use $\Delta t=0.02, 0.01, 0.005$ for mesh~1, mesh~2, and mesh~3, respectively.

\begin{table}
\begin{center}
\begin{tabular}{c|llll}
\hline\\[-1.5ex]
      &$\max\limits_{t\in[0,T]} \|u-u_h\|_{L^2}$ &$(\int_0^T \|\nabla(u-u_h)\|^2 \mathrm{dt})^{\frac12}$ & $(\int_0^T \|u-u_h\|^2\mathrm{dt})^{\frac12}$&$(\int_0^T \|p-p_{h}\|^2\mathrm{dt})^{\frac12}$\\
\hline
mesh~1  &0.37 &0.30 &0.30 &8.10E-2 \\         
mesh~2  &8.60E-2~(2.11) &6.13E-2~(2.29) &5.97E-2~(2.33) &4.04E-2~(1.00) \\
mesh~3  &2.66E-2~(1.69) &1.68E-2~(1.87) &1.62E-2~(1.88) &2.02E-2~(1.00) \\
\hline
\end{tabular}
  \caption{\label{t1} Errors   to analytical solution in $\Omega_{\rm 3D}$ on the sequence of refined meshes. Reduction orders
  are given in brackets}
\end{center}
\end{table}

\begin{table}
\begin{center}
\begin{tabular}{c|ll|ll}
\hline\\[-2ex]
&\multicolumn{2}{c}{in $\Omega_{\rm 1D}^{\rm up}$}&\multicolumn{2}{c}{in $\Omega_{\rm 1D}^{\rm down}$}\\
\hline\\[-1.7ex]
      &$(\int_0^T \|u-u_{h}\|^2\mathrm{dt})^{\frac12}$ &$(\int_0^T \|S-S_{h}\|^2\mathrm{dt})^{\frac12}$
      &$(\int_0^T \|u-u_{h}\|^2\mathrm{dt})^{\frac12}$
      &$(\int_0^T \|p-p_{h}\|^2\mathrm{dt})^{\frac12}$ \\[0.5ex]
\hline
mesh~1    &0.25 &8.84E-4 &0.19 &7.02E-4 \\      
mesh~2    &5.78E-2~(2.11) &2.03E-4~(2.12) &3.67E-2~(2.37) &1.00E-4~(2.81) \\
mesh~3    &1.43E-2~(2.02) &5.01E-5~(2.02) &5.64E-3~(2.70) &2.95E-5~(1.76) \\
\hline
\end{tabular}
\caption{\label{t2} Errors to analytical solution in $\Omega_{\rm 1D}^{\rm up}$ and $\Omega_{\rm 1D}^{\rm down}$ on the sequence of refined meshes. Reduction orders
  are given in brackets}
\end{center}
\end{table}

Based on the energy balance \eqref{energy_balance}, the natural norms for measuring error in $\Omega_{\rm 3D}$  are
$C(0,T,L^2(\Omega_{\rm 3D}))$ and $L^2(0,T,H^1(\Omega_{\rm 3D}))$ for velocity and
$L^2(0,T,L^2(\Omega_{\rm 3D}))$ for pressure. These norms and, additionally, $L^2(0,T,L^2(\Omega_{\rm 3D}))$ for velocity error are shown in Table~\ref{t1}. The error norms in the 1D domains coupled to
the inflow and the outflow boundaries of $\Omega_{\rm 3D}$ are shown in Table~\ref{t2}. We observe the expected second order of convergence for all variables except for pressure in 3D. We remark that  the integral in time error norms were computed
 approximately using the  quadrature rule:
$$
\int_0^T \|\nabla(u-u_h)\|_{L^2}^2 \mathrm{dt}\approx\Delta t \sum_{n=1}^{N}\|\nabla(u(n\Delta t)-u_h^n)\|_{L^2(\Omega_{\rm 3D})}^2,\quad N=T(\Delta t)^{-1}.
$$
Other integral norms were computed in the same way.

\begin{table}
\begin{center}
\begin{tabular}{c|cccc}
\hline\\[-2ex]
      &$\nu=1$&$\nu=0.1$&$\nu=0.01$&$\nu=0.001$\\
\hline
mesh~1 &10.7&10.7&13.4&14.5\\
mesh~2 &5.59&6.69&8.09&8.99\\
mesh~3 &4.42&4.42&6.50&7.88\\
\hline
\end{tabular}
\end{center}
\caption{\label{t3} Average number of iterations of the preconditioned BiCGstab iterations for different meshes
and viscosity parameters.}
\end{table}

\begin{table}
\begin{center}
\begin{tabular}{c|ccccccc}
\hline\\[-2ex]
&\multicolumn{7}{c}{$\nu$}\\
$\Delta t$  & 1 & 0.5 & 0.1 & 0.05 & 0.01 & 0.005 & 0.001 \\
\hline
0.1& 11.78& 10.78& 14.00 & *   &* &* & *\\
0.05& 7.16 &6.84 & 7.21& 12.00 &* &* & *\\
0.01& 4.28 & 4.45 & 5.65& 6.27 &6.30 &6.63 &  7.14\\
\hline
\end{tabular}
\end{center}
\caption{\label{t4} Average number of iterations of the preconditioned BiCGstab iterations for varying time step and viscosity parameters. The results are shown for mesh 2. }
\end{table}

The average number of iterations of the  BiCGstab method for solving \eqref{sp}, with block-triangular preconditioner
\eqref{btal}, are shown in Tables~\ref{t3} and~\ref{t4}. Table~\ref{t3} shows that the convergence of the preconditioned BiCGstab method depends only slightly on the viscosity parameter and improves when the grid is refined. The latter observation is consistent with $h$-independent eigenvalue bounds in \eqref{est1} and with numerical results reported in \cite{ElmSilWat05BOOK} for steady problems. Such robust behavior with respect to $\nu$ is observed only for sufficiently small values of the time step $\Delta t$. The results in Table~\ref{t4} show that for small $\nu$-s the preconditioned  BiCGstab method fails to converge unless $\Delta t=0.01$.

\subsection{Flow around circular cylinder}
\begin{figure}
\centering
\includegraphics[scale=0.66]{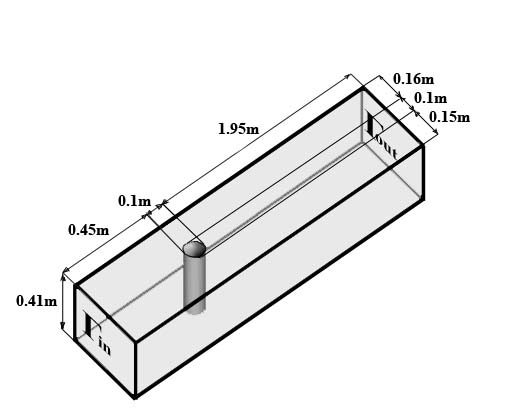}
\caption{\label{fig2b} The  flow domain $\Omega_{\rm 3D}$ in the benchmark problem of the channel flow past a 3D cylinder.}
\end{figure}

Motivated by the simulation of blood flows over an intravenous  filter, we experiment with flows in a 3D domain  having  an inclusion and coupled with  1D model at the outflow boundary. Interesting statistics for such applications are the drag force acting on inclusions and the pressure drop. To validate the ability of the 3D solver to predict these statistics, we
consider two benchmark problems of channel flows past a 3D cylinder with circular cross sections~\cite{SchTur,Braack}.  The 3D flow domain with a cylinder is shown in Figure~\ref{fig2b}.
The no-slip and no-penetration boundary condition $\bu=\mathbf{0}$ is prescribed on the channel walls and the cylinder surface.  The inlet velocity is given by
\[
\bu_{\rm in}=\left(16Uyz(H-y)(H-z)/H^4,0,0\right)^T\quad\mbox{on}~\Gamma_{\rm in},
\]
here $H=0.41m$ is the width of the channel. The kinematic viscosity of the fluid in this test is $\nu=10^{-3}m^2/s$ and its density is $\rho=1kg/m^3$.
 The Reynolds number, $Re= \nu^{-1}D\widetilde{U}$, is defined based on the cylinder width $D=0.1m$ and $\widetilde{U}=\frac23 U$.
We consider the following  two benchmark problems from  \cite{SchTur}:
\begin{itemize}
\item Problem P1: Steady flow with $Re=20$ (${U}=0.45m/s$);
\item Problem P2: Unsteady flow with  varying Reynolds number for ${U}=2.25 \sin(\pi t / 8)m/s$, $t\in[0,8]$.
\end{itemize}
The benchmarks setups do not specify outflow boundary conditions. Hence, on the outflow boundary we apply the 3D-1D coupling using the new conditions  \eqref{cond_b1}, \eqref{newfluxcoupling} so that numerical performance of the coupling can be verified.
\medskip

The statistics  of interest  are the following:
\begin{itemize}
\item The  difference $\Delta p = p(\bx_2) - p(\bx_1)$ between the pressure  values in points $\bx_{1} = \{0.2, 0.205,0.55\}$ and $\bx_{2} = \{0.2,0.205,0.45\}$.
\item The drag coefficient given by an integral over the surface of the cylinder $S \subset \Gamma_{wall}$:
\begin{equation}\label{drag}
C_{\rm drag} = \frac{2}{D H  {U}^2} \int_S \left( \nu \frac{\partial (\bu\cdot\mathbf{t})}{\partial \bn}n_x - p n_z \right) ds
\end{equation}
Here $\bn=(n_x,n_y,n_z)^T$ is the normal vector to the cylinder surface pointing to $\Omega$ and $\mathbf{t}=(-n_z,0,n_x)^T$ is a tangent vector.
\end{itemize}
For problem P2, the reference velocity in \eqref{drag}  is ${U}=2.25m/s$.

\begin{table}\small
\begin{center}
\caption{\label{turek_tou_sou_q1} Problems P1: Computed and reference values of  drag   and pressure drop. }
\begin{tabular}{c|ccccc}
\hline
mesh  & $C_{\rm drag}$ & \%~err & $\Delta p$ & \%~err  & $N_{iter}$\\
\hline
coarse  & 6.149  & 0.58\% & 0.1679 & 1.81\% &   11.5 \\
fine      & 6.196 & 0.17\% & 0.1678 & 1.87\% &  10.5\\
\hline
{Sch\"{a}fer \& Turek} &    [6.05,\,6.25] & & [0.165,\,0.175] &   \\
{Braack \& Richter}    &  6.185           & &   0.1710        &        \\
\hline
\end{tabular}
\end{center}
\end{table}

\begin{table}\small
\begin{center}
\caption{\label{turek_tou_sou_q3} Problems P2: Computed and reference values of  drag   and pressure drop.}
\begin{tabular}{c|cccc}
\hline
mesh  &  $C^{\max}_{\rm drag}$ & \%~err&  $\Delta p(t=8)$& $N_{iter}$ \\
\hline
coarse  &  3.273 & 0.76\%  & -0.115 & 11.7\\
fine      &   3.311 & 0.39\% &  -0.107& 10.6 \\
\hline
{Sch\"{a}fer \& Turek} &   [3.2,3.3]  & &[-0.11, -0.09] \\
{ Bayraktar et al.}     &3.298      &  &--  &   \\
\hline
\end{tabular}
\end{center}
\end{table}

For these benchmark problems, the paper \cite{SchTur} collects several DNS results based on various
finite element, finite volume discretizations of the Navier-Stokes equations and the Lattice Boltzmann method. In \cite{SchTur}, the authors provided reference intervals, where the statistics are expected to converge.
Using a  higher order finite element method and locally refined  adaptive meshes, more accurate
reference values of  $C_{\rm drag}$  and $\Delta p$  were found in~\cite{Braack}
for the steady state solution (problem  P1) and in \cite{BMT12} for unsteady problem P2.
For the computations we use two meshes: a `coarse' and a `fine' ones, both adaptively refined towards
cylinder. The coarser mesh is build of 35803 tetrahedra, which results in 53061 velocity d.o.f. and 8767 pressure d.o.f. for the Taylor-Hood P2-P1
element. The finer  mesh consists of 51634 tetrahedra, which results in 73635
 velocity d.o.f. and 12321  pressure d.o.f.   Both coarse and fine mesh consist of regular tetrahedra. The refinement ratio is about 20 and 60 for the coarse
and the fine meshes, respectively.  We remark that the  fine mesh has four times as many tetrahedra touching the cylinder as the coarse mesh.
The time steps are $\delta t=0.002$ and    $\delta t=0.001$  for the coarse and the fine meshes, respectively.

\begin{figure}[h!]
\centering
\includegraphics[width=0.45\linewidth, trim=100 220 100 220, clip]{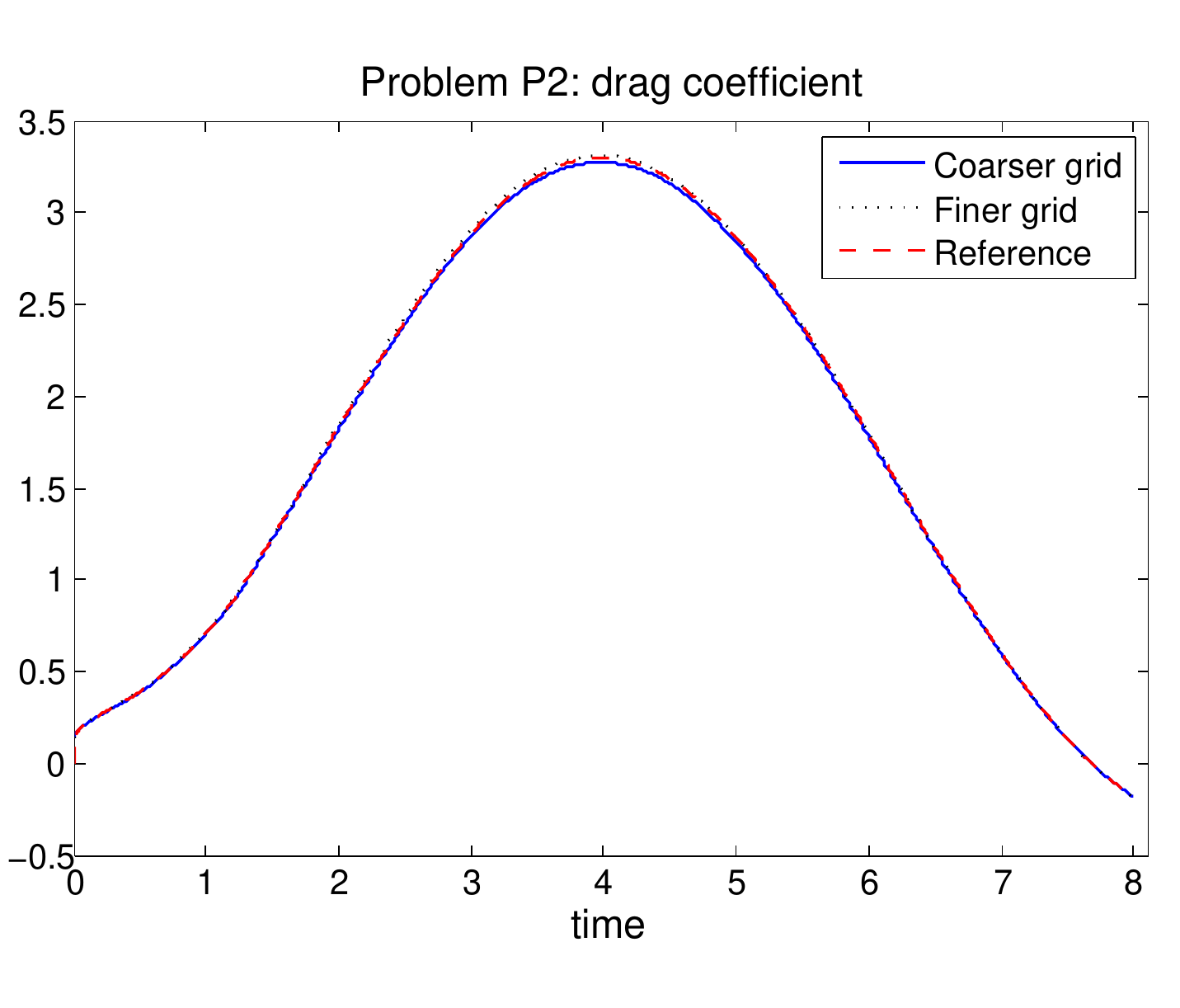}\quad
\includegraphics[width=0.45\linewidth, trim=100 220 100 220, clip]{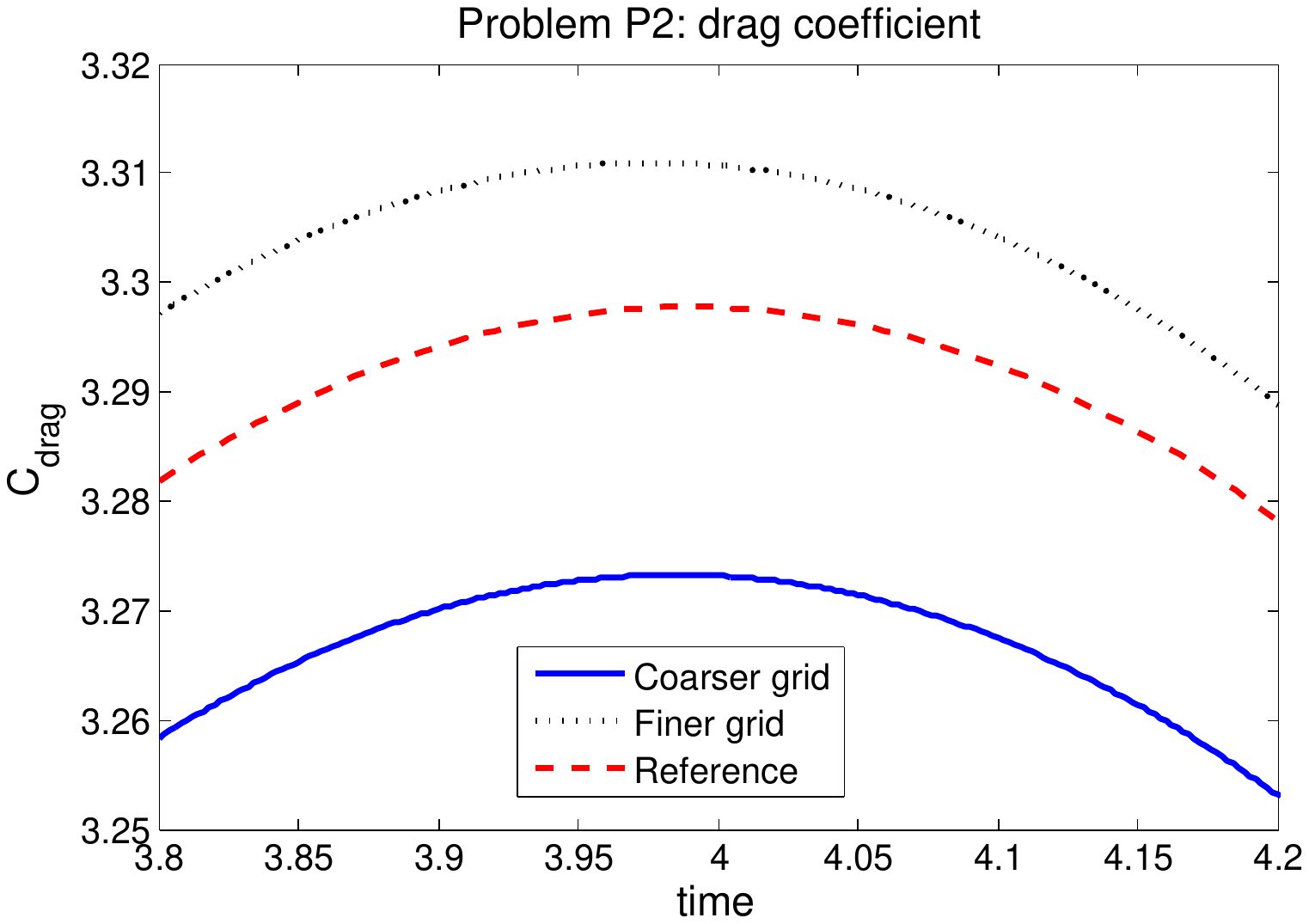}
\caption{\label{fig_drag} Evolution of the drag coefficient for unsteady flow around cylinder: coarse and fine grid results and
reference results. The right figure zooms the plot for time in [3.8,4.2].}\end{figure}

We first show in Tables~\ref{turek_tou_sou_q1} and~\ref{turek_tou_sou_q3} results for problems P1 and P2
obtained with the coarse and the fine meshes. For all settings, the
computed  values are within ``reference intervals'' from  \cite{SchTur} (except $C^{\max}_{\rm drag}$  for problem P2, but in this case the upper reference bound appears to be tough). The computed drag  coefficients were well within 1\% of reference values and pressure drop within 2\%. This is a good result for the number of the degrees of freedom involved.
Indeed, the results shown in  \cite{Braack,BMT12,SchTur} for meshes with about the same number of degrees of freedom show
comparable or worse accuracy.
In Figure~\ref{fig_drag}, we show the computed evolution of the drag coefficient for problem P2  and compare it to the reference results. The computed drag coefficients  match the reference curve very well.
We conclude that the conforming finite element method with the coupling outflow  conditions   is a reliable and stable approach for the simulation of such flow problems.

\section{Simulations of a flow over a model IVC filter} \label{s_filter}

\begin{figure}
\begin{center}
\includegraphics[width=0.33\linewidth]{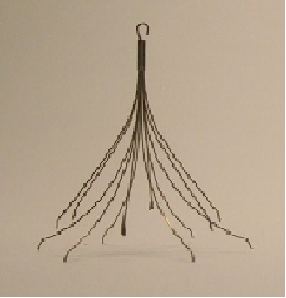}\hskip10ex
\includegraphics[width=0.43\linewidth,trim=80 220 100 210, clip]{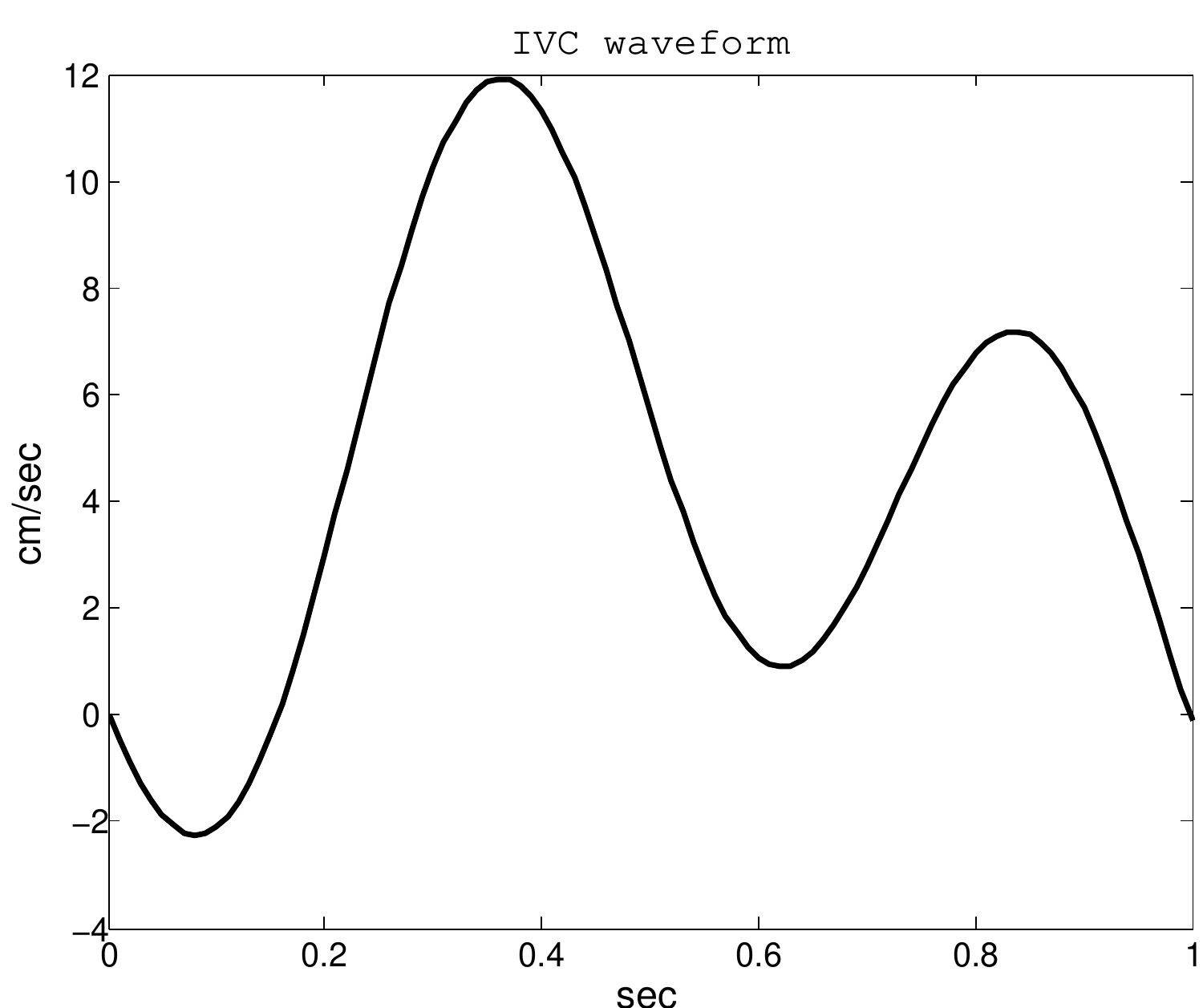}
\end{center}
\caption{\label{figFilt}Left: An example of intravenous filter (Comed Co.); Right: 1D inflow  IVC waveform used in computations. It was designed by interpolating the  IVC Doppler blood flow waveforms from~\cite{cava2}. \label{waveform}}
\end{figure}
The development of endovascular devices  is the challenging problem  of cardiovascular
medicine.  One example is the design of vascular filters implanted in inferior vena cava (IVC) to prevent a blockage of the main artery of the lung or one of its branches
 by a substance that has traveled from elsewhere in the body through the bloodstream. The filter is typically made of thin rigid metal wires as illustrated in Figure~\ref{figFilt} (left).   Numerical simulation is an important  tool that helps in finding an optimal filter design.   Thin and anisotropic construction
of a IVC filter requires adaptive grid refinement  and makes  computations of flows in such domains not an easy task.
In this section, we demonstrate the ability of the numerical  method to treat such problems in a stable way.
One statistic of interest here is the drag force experienced by a filter.  We recall that in this paper we
do not account for the elastic properties of the vessel walls, which are otherwise important in practice.

We consider a segment ($4.5 cm$ long) of IVC  with elliptic cross section $1.6\times2.4 cm$ . The filter is placed on the $0.5cm$ distance from inflow,
it is $2 cm$ long and the diameter of its 12 wire legs is $0.5mm$. Blood is assumed to be
incompressible fluid with dynamic viscosity equal to $0.0055 Pa\times s$ and density equal to $1g/cm^3$.

A blood  flow in IVC  is strongly influenced  by the contraction of the heart. The IVC have pulsatile waveforms
with two peaks and reverse flow~\cite{cava2} occurring on every cardiac cycle.
We consider the Doppler blood flow waveforms of IVC reported in~\cite{cava2} and  approximate them by a smooth periodic function plotted in Figure~\ref{waveform} (right). Note that the presence of significant  reverse flows in IVC differs this problem
from computing arteria flows, where such phenomenon does not typically occur.

On the inflow and outflow,  the 3D vessel is coupled to 1D models as described in section~\ref{s_coupling}.
Each 1D model consists of equations~\eqref{1Dmodel}--\eqref{func_f} posed on intervals of $5cm$ length.
Periodic velocity with waveform as shown in Figure~\ref{waveform} is prescribed on the upstream part of the
1D model coupled to $\Gamma_{\rm in}$.
The maximum 1D model velocity of $12 cm/sec$ yields the maximum inlet velocity in $\Omega_{\rm 3D}$ of about $24cm/sec$. This agrees with the measurements in~\cite{cava2}. The coupling conditions are the same regardless of the mean flow direction.

\begin{figure}
\centering
\includegraphics[width=0.45\linewidth, trim=200 100 200 100, clip]{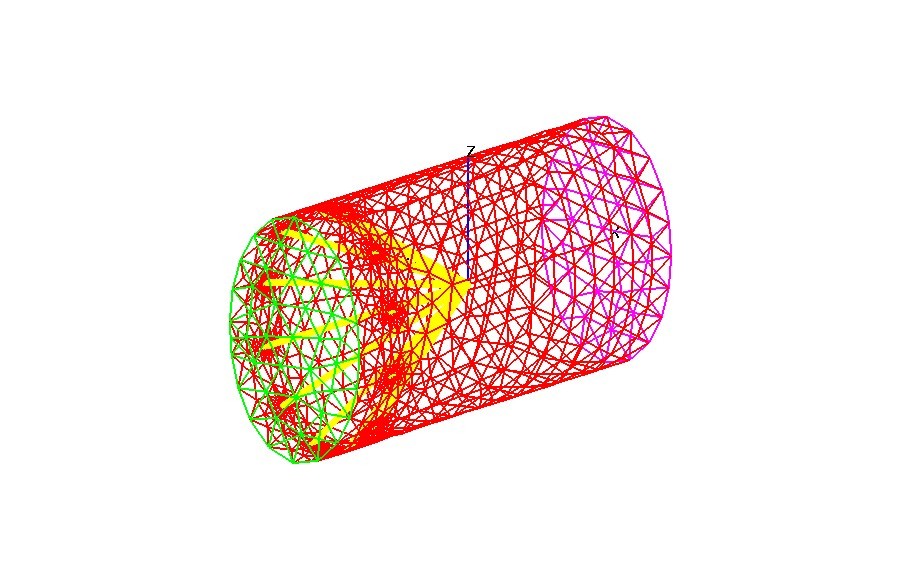}\qquad \includegraphics[width=0.32\linewidth, trim=200 50 200 50, clip]{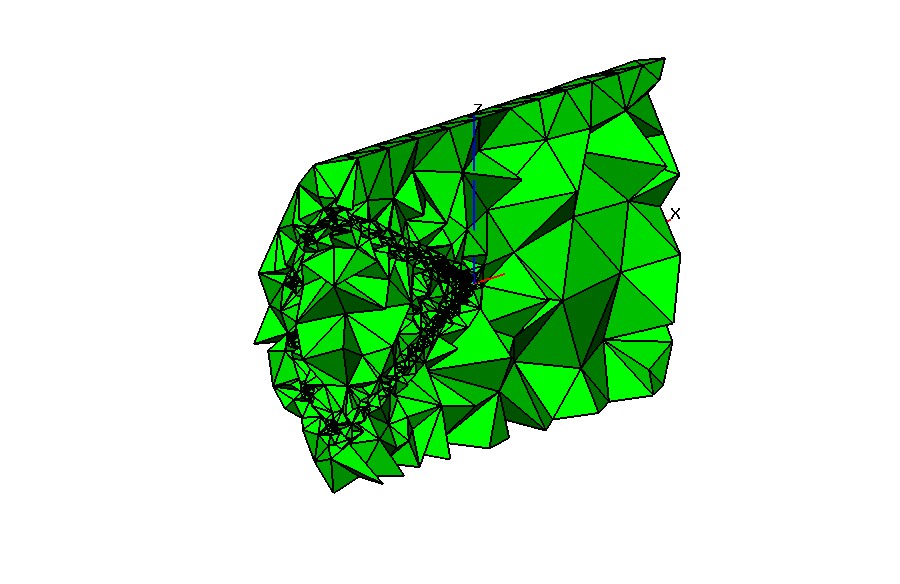} \\
\includegraphics[width=0.8\linewidth]{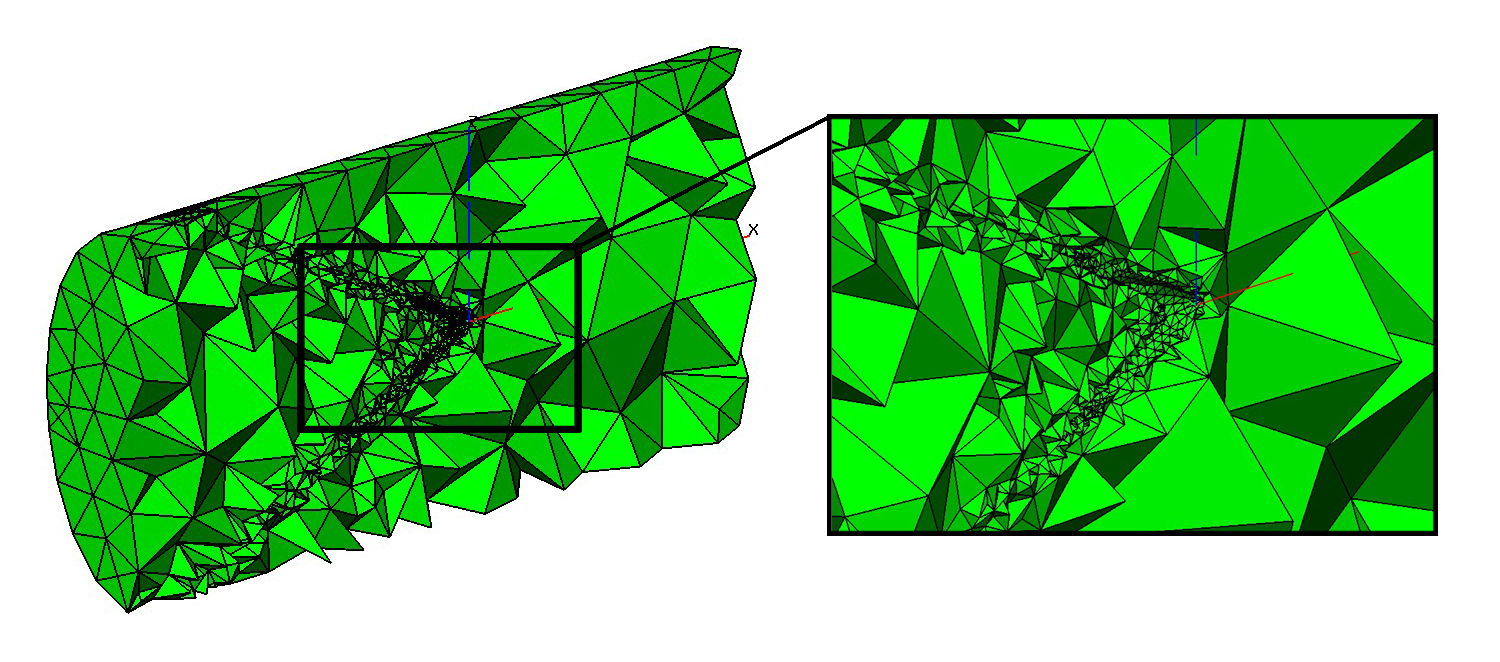}
\caption{\label{fig3} The visualization of the adaptive mesh for the flow over a model IVC filter problem:
the top-left picture shows the boundary surface triangulation; the top-right  picture shows the cutaway views of the tetrahedral grid.  The bottom picture shows the zoom of the mesh in the neighborhood of the filter's `head'. }
\end{figure}

\begin{figure}
\begin{center}
\includegraphics[width=0.47\linewidth]{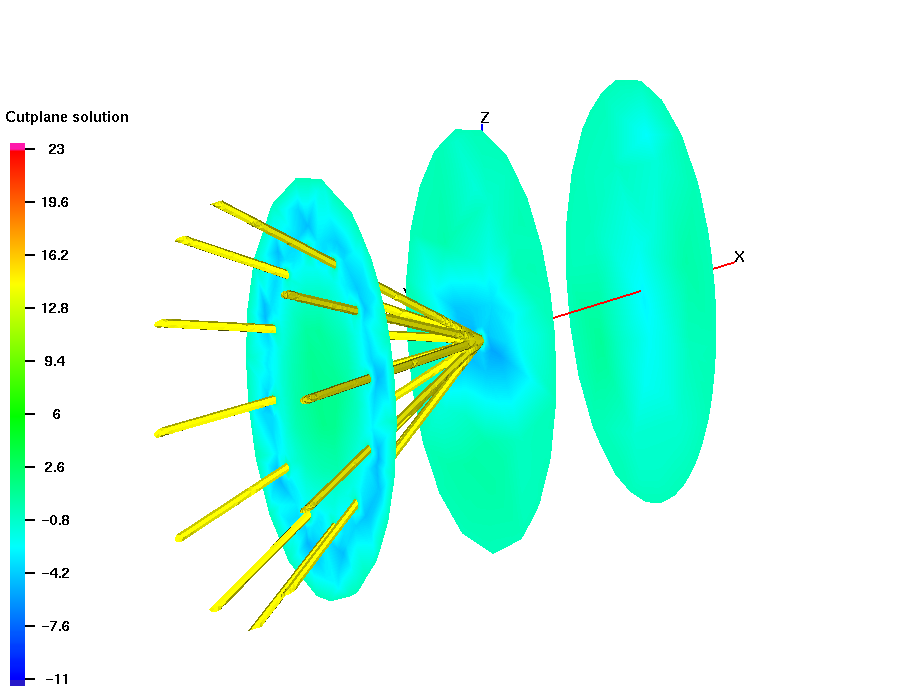}\qquad \includegraphics[width=0.47\linewidth]{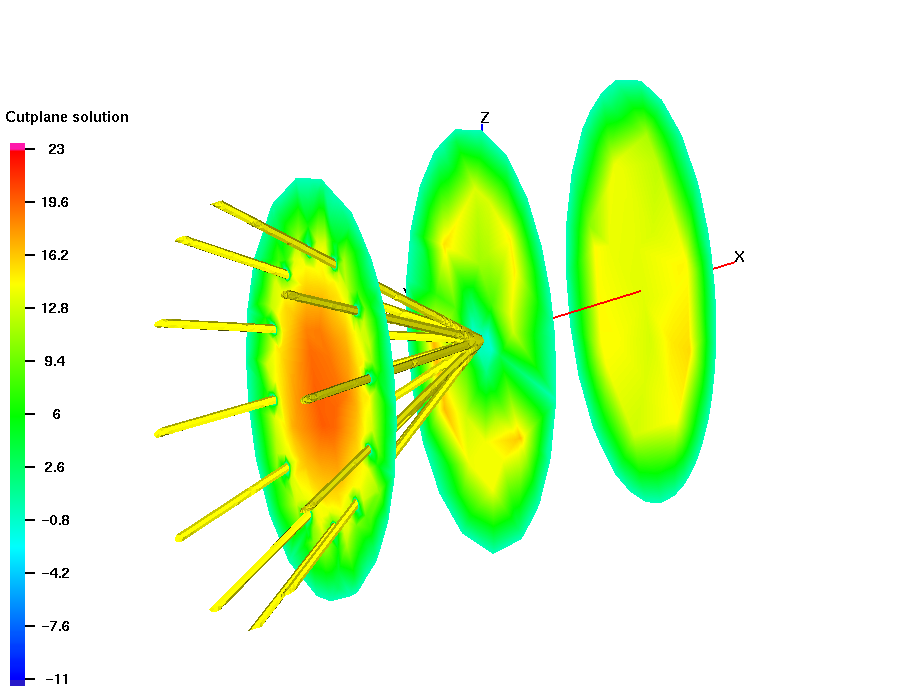}\\\
\includegraphics[width=0.47\linewidth]{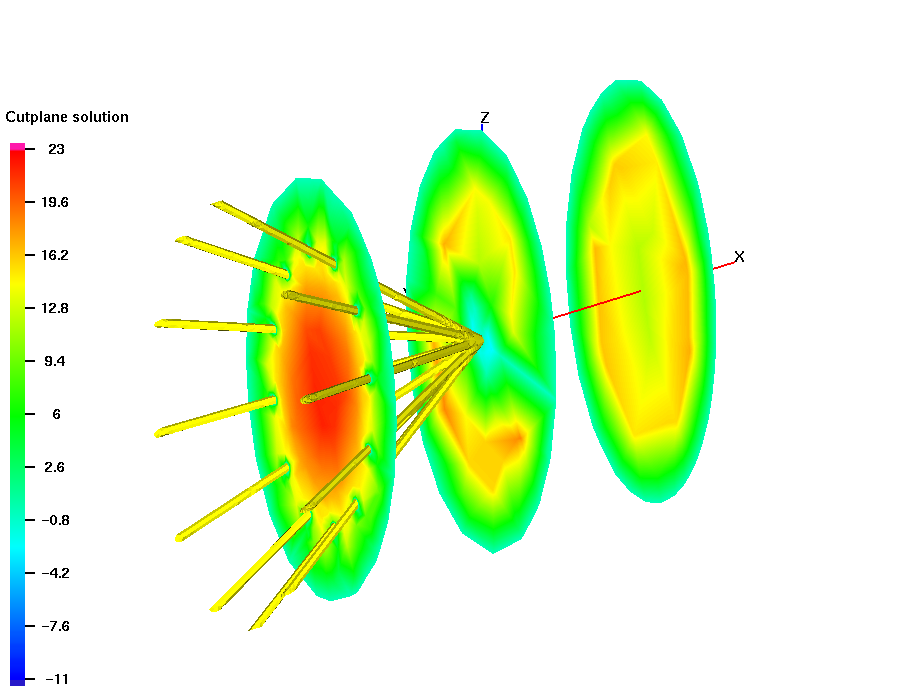}\qquad\includegraphics[width=0.47\linewidth]{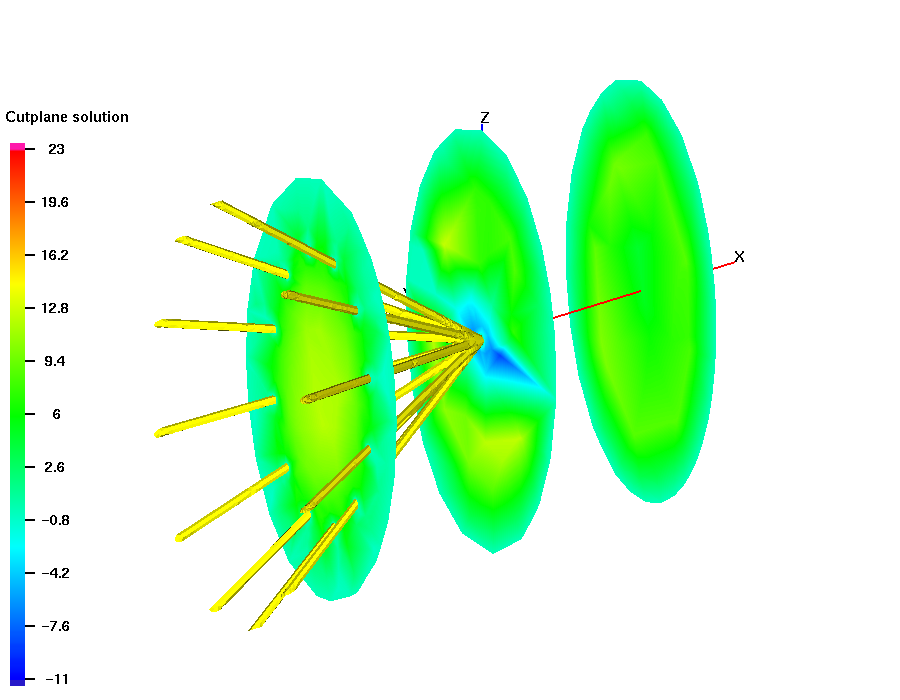}\\
\includegraphics[width=0.47\linewidth]{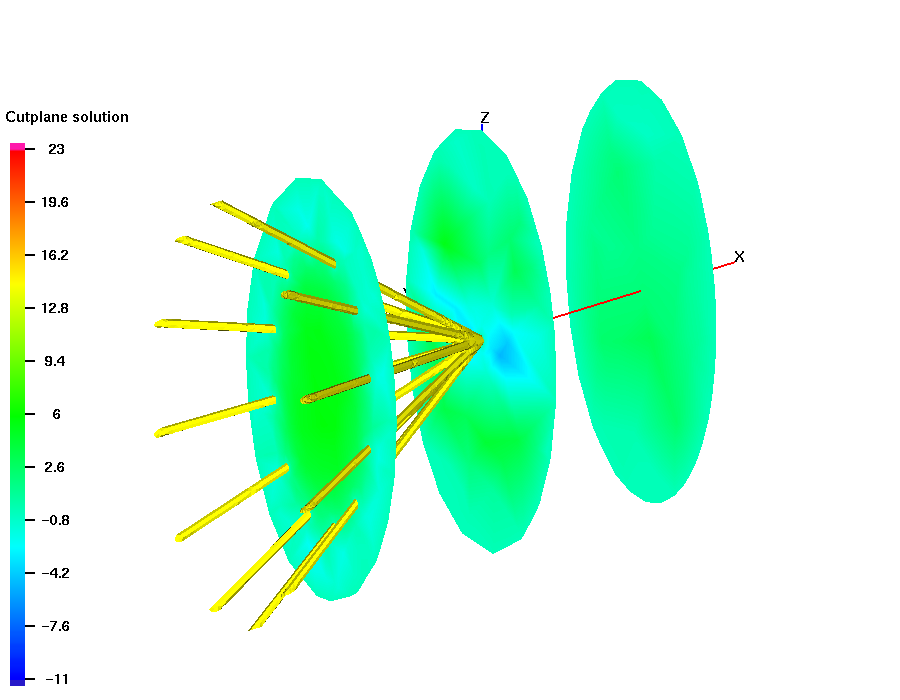}\qquad\includegraphics[width=0.47\linewidth]{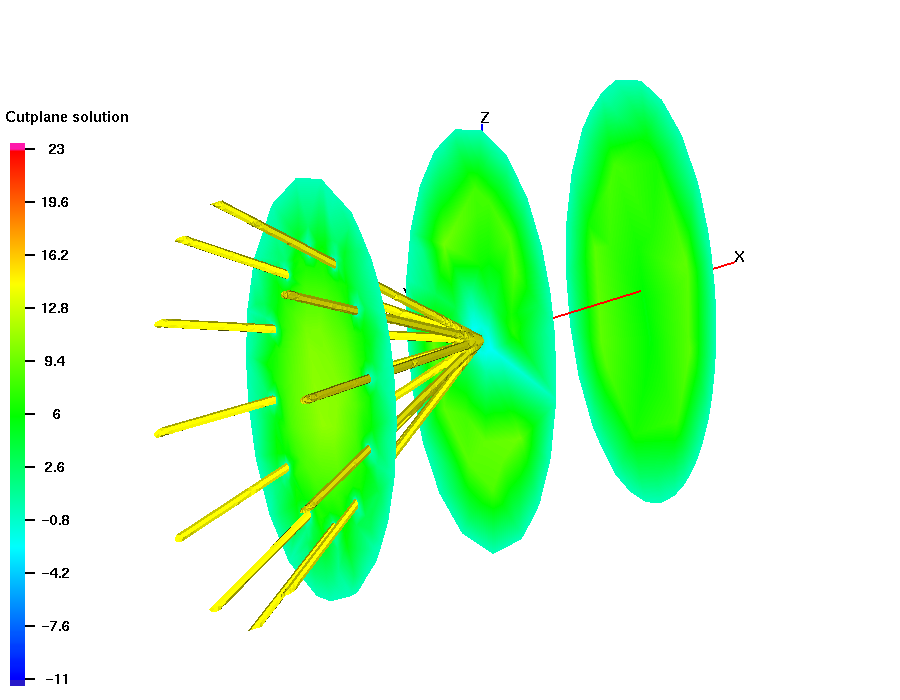}
\end{center}
\caption{\label{fig4} The visualization of  the velocity $x$-component in several cutplanes orthogonal to $x$-axis for
times $t\in\{3.06s,\,3.34s,\,3.39s, 3.52s,3.66s,3.92s\}$. One may note  the occurrence of `returning' flows
behind the filter even for `forward' mean flow.}
\end{figure}

The mesh was  adapted towards the filter, so the ratio of largest and smallest element diameters was about $1.1e+2$, the maximum elements anisotropy
ratio was about $14$. The resulting mesh is illustrated in Figure~\ref{fig3}. The time step in 3D model was set equal
to $0.001sec$. The BiCGstab iterative method, with preconditioner \eqref{btal} was used to solve discrete Oseen subproblems. The stopping criterion was the reduction of the residual by the factor of $10^6$. The average number of linear iterations on every time step was  about 35. We found that choosing  time step larger for this problem, leads to the significant increase of the linear iteration counts and makes `long time' computations non-feasible.

We visualize the computed solutions in Figure~\ref{fig4} by showing the values of the $x$-component of the
velocity in  several cutplanes orthogonal to $x$-axis. Behind the filter the velocity $x$-component eventually has negative values, indicating the occurrence of circulation zones  and `returning' flows. Note that the solution behind the
filter is no longer axial-symmetric: a perturbation to solution induced by non-symmetric tetrahedral grid is sufficient for the von Karman type flow instability to develop behind the filter.

\begin{figure}
\begin{center}
\includegraphics[width=0.47\linewidth,trim=70 210 90 210, clip]{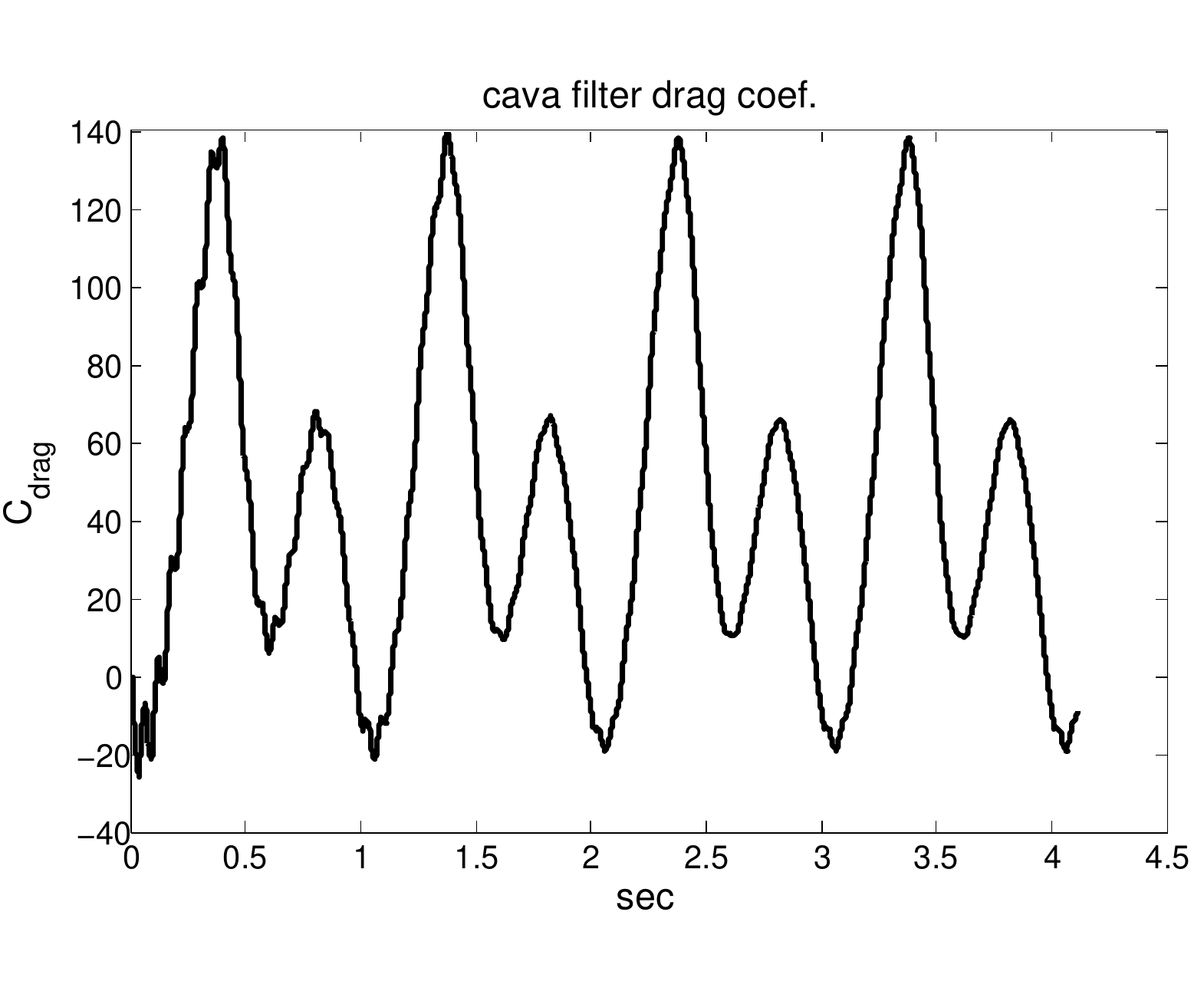}\qquad
\includegraphics[width=0.45\linewidth,trim=75 200 90 210, clip]{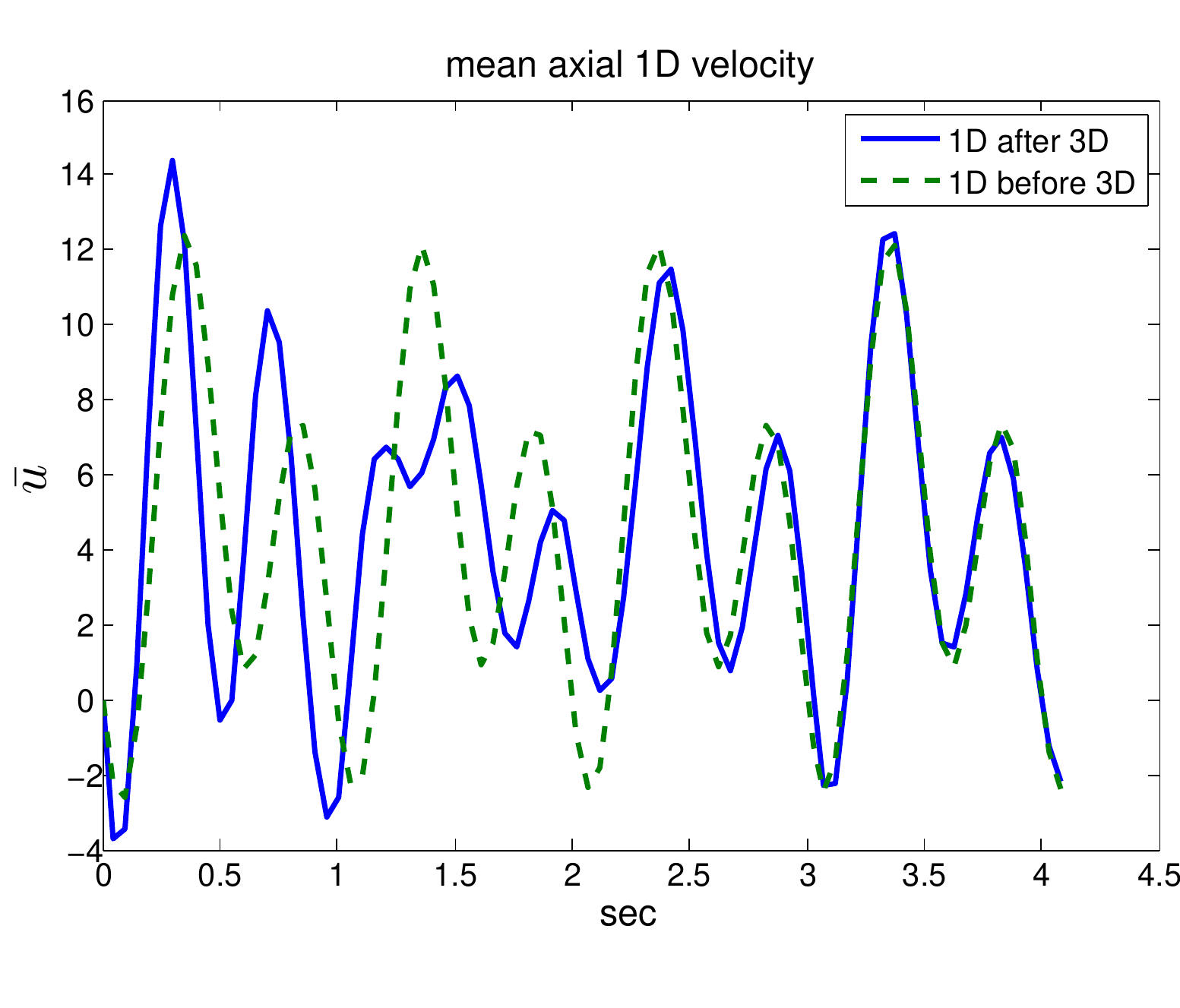}
\end{center}
\caption{\label{fig5} Left: The evolution of the drag force ($g\,sm/s^{2}$) for the IVC filter.
Right: The evolution of the mean axial velocity ($sm/s$) in the middle point of the 1D model \textit{before} and \textit{after} the 3D domain with IVC filter.}
\end{figure}

Figure~\ref{fig5} (left) shows the time evolution of the drag force experienced by the filter. After the instantaneous start, the flow needs few cycles to obtain the periodic regime.     In general, the drag force
follows the pattern of the inflow waveform. In particular, the filter experiences forces both in downstream and upstream directions at different periods of the cardiac cycle.
The right plot in Figure~\ref{fig5} shows the mean axial velocity  in the middle point of the 1D model before and after the 3D domain with cava filter. It is remarkable that after few cycles, when the flow is periodic, the waveforms in the 1D domains
coupled to upstream and downstream boundaries are very close. This suggests that the coupling conditions are efficient in conserving averaged flow quantities such as mean flux.

\section{Conclusions}

We reviewed the 3D and 1D models of fluid flows and some existing coupling conditions for these models.
New coupling conditions were introduced and shown to ensure a suitable bound for the cumulative energy of the model. The conditions were found to perform stable in several numerical tests with analytical and benchmark solutions. For the example of the flow around IVC filter, the coupled numerical model was found to capture the periodic flow regime and correct 1D waveforms before and after 3D domain. The model was  able to handle `opposite direction' flow, i.e. the flow where the `upstream' boundary (boundary with Dirichlet boundary conditions) becomes the outflow boundary for a period of time.
The preconditioned BiCGstab method with one state-of-the-art preconditioner applied to the linearized finite element Navier-Stokes problem performs well. However, often the time step should be taken small enough to make the linear solver converge sufficiently fast.
Overall, the coupled 3D-1D model together with the conforming finite element method and preconditioned iterative strategy was demonstrated as a reliable tool for the simulation of such biological flows as the flow over an inferior vena cava filter.

\section*{Acknowledgements} The authors are grateful to A. Danilov (INM RAS, Moscow) for his help in building tetrahedra meshes in ANI3D, S. Simakov (MIPT, Moscow) for providing us with 1D fluid solver code,
and Yu.Vassilevski (INM RAS, Moscow) for the implementation of the algebraic solver for Problem \eqref{sp}. We would like to thank A. Quaini and S. Canic (UH, Houston, TX) for fruitful discussions and pointing to papers \cite{cava1,cava2}.


\begin{thebibliography}{99}

\bibitem{Abakumov} M.V. Abakumov,  K.V. Gavrilyuk,   N.B. Esikova, A.V. Lukshin, S.I.  Mukhin,  N.V. Sosnin, V.F. Tishkin,  A.P. Favorskij, Mathematical model of the hemodynamics of the cardio-vascular system,  Differ. Equations. 33 (7) (1997) 895--900.

\bibitem{Apel} Th. Apel, H.M. Randrianarivony, Stability of discretizations of the Stokes problem
on anisotropic meshes, Mathematics and Computers in Simulation. 61 (2003) 437--447.

\bibitem{BMT12} E. Bayraktar, O. Mierka,  S. Turek, Benchmark computations of 3D laminar flow around a cylinder with CFX, OpenFOAM and FeatFlow, International Journal of Computational Science and Engineering.  7 (2012), 253--266.

\bibitem{BenziOlshanskii}
{M.~Benzi, M.A.~Olshanskii}, {An augmented Lagrangian-based approach to the Oseen problem},
SIAM J. Sci. Comput. 28 (2006) 2095--2113.

\bibitem{BO11}  M. Benzi,  M.A. Olshanskii, Field-of-values convergence analysis of Augmented Lagrangian
preconditioners for the linearized Navier-Stokes problem, {SIAM J. Numer. Anal.} {49} (2011) 770--788.

\bibitem{BFU} P.J. Blanco, R.A. Feijoro, S.A. Urquiza, A unified variational approach for coupling 3D--1D models
and its blood flow applications, Comput. Methods Appl. Mech. Engrg. 196 (2007) 4391--4410.

\bibitem{BSI} P.J.Blanco, S. Deparis, A.C.I. Malossi,  On the continuity of mean total normal stress in geometrical multiscale cardiovascular problems, EPFL-ARTICLE-182892, 2012

\bibitem{Braack} M. Braack, T. Richter, Solutions of 3D Navier-Stokes benchmark problems with adaptive finite elements, Computers \& Fluids. 35 (2006) 372--392.

\bibitem{BF}  F. Brezzi,  M. Fortin, Mixed and Hybrid Finite Element Methods, Springer, 1991.

\bibitem{CC}
{J.~Cahouet, J. P. Chabard}, {Some fast $3$D finite element solvers for the generalized Stokes problem},
Internat. J. Numer. Methods Fluids. 8 (1988) 869--895.

\bibitem{ChOl}   E.V. Chizhonkov, M.A. Olshanskii, On the domain geometry dependence of the LBB condition, {Math. Modelling Numer. Anal.: $M^2AN$}.  34 (2000) 935--951.

\bibitem{ElmSilWat05BOOK}
{H.C.~Elman, D.J.~Silvester, A.J. Wathen}, {Finite Elements and Fast Iterative Solvers: With Applications in Incompressible Fluid Dynamics}, Numer. Math. Sci. Comput., Oxford University Press, New York, 2005.

\bibitem{Loghin}
{H.C.~Elman, D.~Loghin, A.~J. Wathen},
{Preconditioning techniques for Newton's method for the incompressible Navier--Stokes equations}, BIT. 43 (2003) 961--974.

\bibitem{Tuminaro} H.C. Elman, R.S. Tuminaro,  Boundary conditions in approximate commutator preconditioners for the Navier-Stokes equations, Electronic Transactions on Numerical Analysis. 35 (2009) 257--280.

\bibitem{3D1D1} L. Formaggia, J.F. Gerbeau, F. Nobile, A. Quarteroni, On the coupling of 3D and 1D Navier-Stokes equations for flow problems in compliant vessels, Computer Methods in Applied Mechanics and Engineering. 191 (2001) 561--582.

\bibitem{3D1D2} L. Formaggia, A. Moura, F. Nobile,   On the stability of the coupling of 3D and 1D fluid-structure interaction models for blood flow simulations, ESAIM: Mathematical Modelling and Numerical Analysis. 41 (4) (2007) 743--769.

\bibitem{Garbey}
{M.~Garbey, Yu.~A.~Kuznetsov, Yu.~V.~Vassilevski}, {Parallel Schwarz method for a convection-diffusion problem},
SIAM J. Sci. Comput. 22 (2000) 891--916.

\bibitem{GPS}
{A.~Greenbaum, V.~Pt\'ak, Z.~Strako\u{s}}, {Any nonincreasing convergence curve is possible for GMRES},
SIAM J. Matrix Anal. Appl. 17 (1996) 465--469.

\bibitem{HR97} J.G. Heywood, R. Rannacher, S. Turek, Artificial boundaries and flux and pressure conditions for the incompressible Navier-Stokes equations, International Journal for Numerical Methods in Fluids. 22 (1996) 325--352.

\bibitem{KayLogWat02} {D.~Kay, D.~Loghin,  A.J.~Wathen}, {A preconditioner for the steady-state Navier--Stokes equations}, SIAM J. Sci. Comput. 24 (2002) 237--256.

\bibitem{KS} {A.~Klawonn, G.~Starke}, {Block triangular preconditioners for nonsymmetric saddle point problem},
Numer. Math. 81 (1999) 577--594.

\bibitem{LMNOR} {W. Layton, C.C. Manica, M. Neda, M.A. Olshanskii, L.G. Rebholz}, On the accuracy of the rotation form in simulations of the Navier-Stokes equations, {Journal of Computational Physics}. {228} (2009) 3433--3447.

\bibitem{O2000} M.A.  Olshanskii, A low order Galerkin finite element method for the Navier-Stokes
equations of steady incompressible flow: A stabilization issue and
iterative methods, {Comp. Meth. Appl. Mech. Eng.} {191} (2002) 5515--5536.

\bibitem{OR06} {M.A. Olshanskii, J. Peters, A. Reusken}, Uniform preconditioners for a
parameter dependent saddle point problem with application to
generalized Stokes interface equations, {Numerische Mathematik}. {105} (2006) 159--191.

\bibitem{OV07}  M.A. Olshanskii,  Yu.V. Vassilevski, Pressure Schur complement preconditioners for the discrete Oseen problem, SIAM J.Sci.Comp.  {29} (2007) 2686--2704.

\bibitem{Papa} G. Papadakis, Coupling 3D and 1D fluid�structure-interaction models for wave
propagation in flexible vessels using a finite volume pressure-correction scheme, Commun. Numer. Meth. Engng. 25 (2009) 533--551.

\bibitem{cava1} W. T. Pua, T. Ishiwatac, A. L. Juraszeka, Q. Mac, S. Izumo, GATA4 is a dosage-sensitive regulator of cardiac morphogenesis, Developmental Biology. 275 (2004) 235--244.


\bibitem{Simakov} S.S. Simakov,  A.S. Kholodov, Computational study of oxygen concentration in human blood under low frequency disturbances, Mathematical models and computer simulations. 1 (2) 2009 283--295.


\bibitem{SchTur}  M. Sch\"{a}fer, S. Turek, The benchmark problem ``Flow around a cylinder''. In Hirschel EH (ed.), Flow Simulation with High-Performance Computers II, vol. 52. Notes on Numerical Fluid Mechanics, Vieweg. 1996; 547--566

\bibitem{Stuben}   J.W. Ruge, K. St\"{u}ben, Algebraic multigrid, in Multigrid Methods, edited by S. F. McCormick, p. 73 (SIAM, Philadelphia, PA, 1987).

\bibitem{Vally} A. Quarteroni, A. Valli, {Domain Decomposition Methods for Partial Differential Equations},
Oxford University Press, Oxford, UK, 1999.

\bibitem{QTV} A. Quarteroni, L. Formaggiaa, A. Veneziani, Cardiovascular Mathematics: Modeling and Simulation of the Circulatory System, Springer-Verlag Itali, Milano 2009

\bibitem{UBVF}    S.A. Urquiza, P.J. Blanco, M.J. Vernere, R.A. Feijoro, Multidimensional modelling for the carotid artery blood flow, Comput. Methods Appl. Mech. Engrg. 195 (2006) 4002--4017.

\bibitem{VSK} Y.V. Vassilevski, S.S. Simakov, S.A. Kapranov, A multi-model approach to intravenous filter optimization, International Journal for Numerical Methods in Biomedical Engineering. 26 (2010) 915--925.

\bibitem{all1} Yu. Vassilevskii, S. Simakov, V. Salamatova, Yu. Ivanov, T. Dobroserdova, Numerical issues of modelling blood flow in networks of vessels with pathologies, Russian Journal of Numerical Analysis and Mathematical
 Modelling. 26 (6) (2011) 605--622.

\bibitem{all2} Y. Vassilevski, S. Simakov, V. Salamatova, Y. Ivanov, T. Dobroserdova, Vessel wall models for simulation of atherosclerotic vascular networks,  Mathematical Modelling of Natural Phenomena. 6 (7) (2011) 82--99.

\bibitem{all3} Y. Vassilevski, S. Simakov, V. Salamatova, Y. Ivanov, T. Dobroserdova, Blood flow simulation in atherosclerotic vascular network using fiber-spring representation of diseased wall, Mathematical Modelling of Natural Phenomena. 6 (5) (2011) 333--349.

\bibitem{Vignon} I.E. Vignon-Clementel, C.A. Figueroa, K.E. Jansen, C.A. Taylor, Outflow boundary conditions for three-dimensional finite element modeling of blood flow and pressure in arteries, Comput. Methods Appl. Mech. Engrg. 195 (2006) 3776--3796.

\bibitem{cava2} D. Zhang, T. Kanzaki, Doppler waveforms: the relation between ductus
venosus and inferior vena cava, Ultrasound in Med. \& Biol. 31 (9) (2005)  1173--1176.

\bibitem{ani3D}  ANI3D: Advanced Numerical Instruments. http://sourceforge.net/projects/ani3d

\end{thebibliography}
\end{document}